\documentclass[aps,amsmath,amssymb,preprint,prd,showpacs,showkeys]{revtex4-1}
\usepackage[b]{esvect}
\usepackage{graphicx}
\usepackage{dcolumn}

\usepackage{bm}
\usepackage{makecell}
\usepackage{footnote}
\usepackage{multirow}
\usepackage{xcolor}
\usepackage[colorlinks=true]{hyperref}
\usepackage{axodraw}
\usepackage{subcaption}
\begin{document}
\title{Form Factors and Decay of $\bar{B}_s^0\to J/\psi \phi$ From QCD Sum Rule}

\author{Ying-Quan Peng}
\email{2120170119@mail.nankai.edu.cn}
\author{Mao-Zhi Yang}
\email{yangmz@nankai.edu.cn}
\affiliation{School of Physics, Nankai University,Tianjin 300071, People's Republic of China}

\begin{abstract}
 We calculate $\bar{B}_s^0\to \phi$ translation form factors $V$, $A_0$, $A_1$, $A_2$ based on QCD sum rule and study the nonleptonic two-body decay of $\bar{B}_s^0\to J/\psi \phi$ with the form factors obtained. We calculate the time-integrated branching ratio of $\bar{B}_s^0\to J/\psi \phi$ decay. The results for both the total branching ratio and the cases for the final vectors in longitudinal and transverse polarizations are consistent with experimental data.
\end{abstract}


\maketitle

\section{Introduction}
$CP$ asymmetry arises due to the non-vanishing complex phase of the Cabibbo-Kobayashi-Maskawa (CKM) matrix elements \cite{KM,Cabibbo} in the standard model (SM). The requirement of unitarity of the CKM matrix results in a set of triangles in the complex plane. $\bar{B}_s^0(t)\to J/\psi \phi$ is a golden decay mode to measure the angle $\beta_s$, one of the angles of the triangle in the $bs$ sector: $V_{us}V_{ub}^*+V_{cs}V_{cb}^*+V_{ts}V_{tb}^*=0$. The angle $\beta_s$ is related to the sides of unitarity triangle by $\beta_s=arg[-V_{ts}V_{tb}^*/(V_{cs}V_{cb}^*)$ \cite{R17}. $\bar{B}_s(t)\to J/\psi \phi$ decay stimulate wide interest in both theory and experiment \cite{MA2016,CMS1,Lei2010,Colan2011,Liu2014}. This decay mode is not only interesting for analysis of $CP$ violation, but also useful for studying strong interaction in the decay process. Factorization is a basic method to calculate non-leptonic $B$ meson decays, where it is assumed that the hadronic decay amplitude can be factorized as a product of matrix elements of two local quark-antiquark currents \cite{Fak1978,Cab1978-1}. The QCD non-factorizable corrections to the factorization result can be calculated systematically in perturbation theory, which is called QCD factorization \cite{Ben1999,Ben2000}. The hadronic matrix element of the quark-antiquark current in $B$ decays can be decomposed as polynomials of form factors. The form factors are in general non-perturbative quantities in QCD, which can be calculated with non-perturbative method, such as Lattice QCD, QCD sum rule, QCD light-cone sum rule, and quark model, etc. The $\Bar{B}^0_s\to\phi$ transition form factors involved in $\bar{B}_s^0\to J/\psi\phi$ decay have been calculated by QCD light-cone sum rule (LCSR) \cite{Ball1998,R11}, Quark model (QM) \cite{R12}, and QCD sum rule \cite{R15} in literature. Some form factors obtained by QCD sum rule in Ref. \cite{R15} are different from other results calculated in Refs. \cite{R11,R12} by a sign. Here in this work, we revisit the $\Bar{B}^0_s\to\phi$ transition form factors with QCD sum rule method. Then the from factors obtained in this work are used to  study the nonleptonic decay of $\bar{B}_s^0\to J/\psi \phi$. The the transverse, longitudinal and total time-dependent decay widths $\Gamma_L(t)$, $\Gamma_T(t)$ and $\Gamma(t)$ are calculated respectively. The results are consistent with experimental data within experimental and theoretical uncertainties.

The remainder of this paper is organized as follows: In sec. II, we present the method to calculate the form factors in QCD sum rule method. Section III is for the numerical analysis, and Section IV for the application of the form factors in the decay mode $\bar{B}_s^0\to J/\psi \phi$. Section V is a brief summary.


\section{Theoretical Framework}
\subsection*{1. The method}

 In the factorization approach, one ingredient of the amplitude of the decay mode $\bar{B}_s^0\to J/\psi \phi$ is the hadronic matrix element
 $\langle \phi|\bar{s}\gamma_\nu (1-\gamma_5)b|\bar{B}_s^0 \rangle$, which can be decomposed as
\begin{eqnarray}\label{10}
\langle\phi(\varepsilon,p_2)|\bar{s}\gamma_\nu (1-\gamma_5)b|\bar{B}_s^0(p_1)\rangle
&=&\varepsilon_{\nu\rho\alpha\beta}\varepsilon^{*\rho}p_1^\alpha
p_2^\beta
\frac{2V(q^2)}{m_{B_s}+m_{\phi}} \nonumber \\
&&-i(\varepsilon^*_\nu-\frac{\varepsilon^*\cdot q}{q^2}
q_\nu)(m_{B_s}+m_{\phi})A_1(q^2) \\
&&+i[(p_1+p_2)_\nu -\frac{m_{B_s}^2-m_{\phi}^2}{q^2} q_\nu ]
\varepsilon^*\cdot q \frac{A_2(q^2)}{m_{B_s}+m_{\phi}} \nonumber\\
&&-i\frac{2m_\phi \varepsilon^*\cdot q}{q^2}q_\nu
A_0(q^2),\nonumber
\end{eqnarray}
where the parameters $V$, $A_0$, $A_1$, $A_2$ are the transition form factors, and $q=p_1-p_2$.

The QCD sum rule method was originally developed by Shifman, Vainshtein and Zakharov in the late 1970s \cite{SVZ1,SVZ2}, which was widely used in hadronic process, see Ref. \cite{Colan2000} for a review.
In order to calculate the transition form factors of $\bar{B}^0_s$ to $\phi$ in QCD, we consider the three-point correlation function defined by
\begin{equation}\label{11}
\Pi_{\mu\nu}=i^2\int d^4 x d^4 y e^{ip_2\cdot x-ip_1\cdot y}
 \langle 0|T\{j^\phi_\mu(x) j_\nu (0)j_5(y)\} |0\rangle ,
\end{equation}
where the three currents are: (1) $j_5(y)=\bar{b}(y)i\gamma_5 s(y)$, the current of $\bar{B}_s^0$ channel;
(2) $j_\nu (0)=\bar{s}\gamma_\nu (1-\gamma_5)b$, the current of weak transition; (3) $j_\mu^\phi(x)=\bar{s}(x)\gamma_\mu s(x)$, the current of $\phi$ channel.

One can use the double dispersion relation to express the correlation function as
\begin{equation}\label{12}
\Pi_{\mu\nu}=\int ds_1 ds_2\frac{\rho (s_1,s_2,q^2)}{(s_1-p_1^2) (s_2-p_2^2)},
\end{equation}
where $\rho (s_1,s_2,q^2)$ is the spectral density. By inserting a full set of intermediate hadronic states into the time-ordered product in the correlation function, one can obtain the spectral density function as
\begin{equation}\label{rho}
\rho (s_1,s_2,q^2)=\sum_{X}\sum_{Y}\langle 0|j_\mu^\phi |X\rangle \langle
   X|j_\nu |Y\rangle \langle Y|j_5 |0 \rangle \delta (s_1-m_Y^2)
    \delta (s_2-m_X^2)\theta(p_X^0)\theta(p_Y^0),
\end{equation}
where $X$ and $Y$ denote the full set of hadronic states of
$\phi$ and $\bar{B}^0_s$ channels, respectively. Substituting the spectral density $\rho (s_1,s_2,q^2)$ in Eq. (\ref{rho}) into Eq.~(\ref{12}) and integrating over $s_1$ and $s_2$, then we can get
\begin{equation}\label{13}
\Pi_{\mu\nu}=\sum_{X}\sum_{Y}\frac{\langle 0|j_\mu^\phi |X\rangle \langle
   X|j_\nu |Y\rangle \langle Y|j_5 |0
   \rangle}{(m_Y^2-p_1^2)(m_X^2-p_2^2)} +\mbox{\small{continuum states}}.
\end{equation}
The result of the correlation function in the above equation can be also expressed as a sum of
the ground states, excited states and continuum states
\begin{equation}\label{14}
\Pi_{\mu\nu}=\frac{\langle 0|j_\mu^\phi |\phi\rangle \langle
   \phi|j_\nu |\bar{B}_s^0\rangle \langle \bar{B}_s^0|j_5 |0
   \rangle}{(m_{B_s}^2-p_1^2)(m_{\phi}^2-p_2^2)} +
   \mbox{\small{excited states~+~continuum states}}.
\end{equation}
Using the definition of decay constants in the following
 \begin{eqnarray}\label{15}
&&\langle 0|\bar{s}\gamma_\mu s|\phi\rangle =m_\phi f_\phi
  \varepsilon_\mu^{(\lambda)}, \nonumber \\
&&\langle 0|\bar{s}\gamma_\mu\gamma_5 s|\phi\rangle =0, \nonumber \\
&&\langle 0|\bar{s}i\gamma_5 b|\bar{B}_s^0\rangle =
\frac{f_{B_s}m_{B_s}^2}{m_b+m_s},
\end{eqnarray}
where $f_\phi$ and $f_{B_s}$ are decay constants of the relevant mesons, the correlation function is changed to be
\begin{eqnarray}
\Pi_{\mu\nu}=\frac{ m_\phi f_\phi
  \varepsilon_\mu^{(\lambda)}\langle
   \phi(\varepsilon_\mu^{(\lambda)},p_2)|j_\nu |\bar{B}_s^0(p_1)\rangle
   f_{B_s} m_{B_s}^2}{(m_{B_s}^2-p_1^2)(m_{\phi}^2-p_2^2)(m_b+m_s)}\nonumber\\
    +   \mbox{\small{excited and continuum states}}.
    \label{16}
\end{eqnarray}

Meanwhile the time-ordered current operator in the correlation function in Eq.~(\ref{11}) can be expanded in terms of a series of local operators with increasing dimensions in QCD
\begin{eqnarray}
&&i^2\int d^4x d^4 y e^{ip_2\cdot x-ip_1\cdot y}
 T\{j^\phi_\mu(x) j_\nu (0)j_5(y)\} \nonumber\\
 &=&C_{0\mu\nu} I +C_{3\mu\nu} \bar{\Psi}\Psi
    +C_{4\mu\nu} G^a_{\alpha\beta}G^{a\alpha\beta}
    +C_{5\mu\nu} \bar{\Psi}\sigma_{\alpha\beta}T^a G^{a\alpha\beta}\Psi
    \nonumber\\
  &~+&C_{6\mu\nu}
 \bar{\Psi}\Gamma \Psi \bar{\Psi}\Gamma^{\prime}\Psi+\cdots,
 \label{17}
 \end{eqnarray}
where $C_{i\mu\nu}$ are Wilson coefficients, $I$ the unit
operator, $\bar{\Psi}\Psi$ the local fermion field operator of
light quarks, $G^a_{\alpha\beta}$ gluon strength tensor,
$\Gamma$ and $\Gamma^{\prime}$ the matrices appearing in the
procedure of calculating the Wilson coefficients. At deep negative values of $p_1^2$ and $p_2^2$, the Wilson coefficients can be calculated reliably in perturbative QCD, and the operator-product expansion (OPE) in Eq. (\ref{17}) can converge quickly.

Considering the non-vanishing vacuum-expectation-value of the operators in Eq.~(\ref{17}), we can get the correlation function in terms of Wilson coefficients and condensates of local operators
\begin{eqnarray}
\Pi_{\mu\nu}&=&i^2\int d^4x d^4 y e^{ip_2\cdot x-ip_1\cdot y}
 \langle 0|T\{j^\phi_\mu(x) j_\nu (0)j_5(y)\}|0\rangle \nonumber\\
 &=&C_{0\mu\nu} I +C_{3\mu\nu} \langle 0|\bar{\Psi}\Psi|0\rangle
    +C_{4\mu\nu} \langle 0|G^a_{\alpha\beta}G^{a\alpha\beta}|0\rangle
    +C_{5\mu\nu} \langle 0|\bar{\Psi}\sigma_{\alpha\beta}T^a G^{a\alpha\beta}\Psi|0\rangle
    \nonumber\\
  &~+&C_{6\mu\nu}\langle 0|
 \bar{\Psi}\Gamma \Psi \bar{\Psi}\Gamma^{\prime}\Psi|0\rangle +\cdots,
 \label{18}
\end{eqnarray}
According to the Lorentz structure of the correlation function, Eq.~(\ref{18}) can be re-expressed by six parts
\begin{equation}
\Pi_{\mu\nu}=f_0\varepsilon_{\mu\nu\alpha\beta}p_1^\alpha
p_2^\beta-i(f_1 p_{1\mu}p_{1\nu}+f_2 p_{2\mu}p_{2\nu}+f_3
p_{1\mu}p_{2\nu}+f_4 p_{1\nu}p_{2\mu}+f_5g_{\mu\nu}). \label{19}
\end{equation}
The coefficients $f_i$'s are consisted of perturbative and condensate contributions,
\begin{equation}\label{20}
f_i=f_i^{pert}+f_i^{(3)}+f_i^{(4)}+f_i^{(5)}+f_i^{(6)}+\cdots,
\end{equation}
where $f_i^{pert}$ is the perturbative contribution of the
unit operator, and $f_i^{(3)}$, $f_i^{(4)}$, $f_i^{(5)}$, $f_i^{(6)}$, $\cdots$, are
contributions of condensates of operators with increasing dimension
in OPE.

In next section we shall know that perturbative
contribution and gluon-condensate contribution can be written in the form of dispersion integration
\begin{eqnarray}
 f_i^{pert}&=&\int d s_1
 d s_2\frac{\rho^{pert}_i(s_1,s_2,q^2)}{(s_1-p_1^2)(s_2-p_2^2)},
 \nonumber  \\
f_i^{(4)}&=&\int d s_1
 d s_2\frac{\rho^{(4)}_i(s_1,s_2,q^2)}{(s_1-p_1^2)(s_2-p_2^2)}.
 \nonumber  \nonumber
 \end{eqnarray}
We can approximate the contribution of excited states and continuum states as
integrations over some thresholds $s_1^0$ and $s_2^0$ in the above
two equations. Then equating the two expressions of the correlation function in
Eq.~(\ref{16}) and (\ref{19}), we can get an equation for extracting the form factors. But such
an equation may heavily depend on the approximation
for the contribution of excited states and the contributions of
higher dimensional operators in OPE. To improve such an equation and make the contribution
of higher dimensional operator small, one
can make Borel transformation over $p_1^2$ and $p_2^2$ in both
sides, which can suppress the contributions of excited states and
condensate of higher dimensional operators. The definition of
Borel transformation to any function $f(x^2)$ is
$$\hat{B}_{\left|\frac{}{}\right.x^2,M^2}f(x^2)=\lim_{\small\begin{array}{ll}& k\to\infty, x^2\to -\infty  \\&-x^2/k=
    M^2  \end{array} } \frac{(-x^2)^k}{(k-1)!}\frac{\partial ^k}{\partial (x^2)^k}
    f(x^2).$$

Matching these two expressions of the correlation function in
Eq.~(\ref{16}) and (\ref{19}), and performing
Borel transformation for both variables $p_1^2$ and $p_2^2$, the sum rules for the form factors
can be obtained
\begin{eqnarray}
V(q^2)&=& -\frac{(m_b+m_s)(m_{B_s}+m_\phi )}{2m_\phi f_\phi
 f_{B_s}m_{B_s}^2}e^{m_{B_s}^2/M_1^2}e^{m_{\phi}^2/M_2^2}M_1^2M_2^2
 \hat{B}f_0, \nonumber \\
A_1(q^2)&=& -\frac{(m_b+m_s)}{m_\phi f_\phi
 f_{B_s}m_{B_s}^2(m_{B_s}+m_\phi )}e^{m_{B_s}^2/M_1^2}e^{m_{\phi}^2/M_2^2}M_1^2M_2^2
 \hat{B}f_5, \nonumber \\
A_2(q^2)&=& \frac{(m_b+m_s)(m_{B_s}+m_\phi )}{m_\phi f_\phi
 f_{B_s}m_{B_s}^2}e^{m_{B_s}^2/M_1^2}e^{m_{\phi}^2/M_2^2}M_1^2M_2^2
 \frac{1}{2}\hat{B}(f_1+f_3), \label{21} \\
A_0(q^2)&=& -\frac{(m_b+m_s)}{2m_\phi^2 f_\phi
 f_{B_s}m_{B_s}^2}e^{m_{B_s}^2/M_1^2}e^{m_{\phi}^2/M_2^2}M_1^2M_2^2
 [\hat{B}(f_1+f_3)\frac{m_{B_s}^2-m_\phi^2}{2}\nonumber \\&~&+
 \hat{B}(f_1-f_3)\frac{q^2}{2}+\hat{B}f_5], \nonumber
 \end{eqnarray}
where $\hat{B} f_i$ denotes Borel transformation of $f_i$ for both
variables $p_1^2$ and $p_2^2$. $M_1$ and $M_2$ are Borel
parameters. After subtracting the contribution of the excited
states and continuum states, the dispersion integration for
perturbative and gluon condensate contribution should be performed
under the threshold
\begin{eqnarray}
 f_i^{pert}&=&\int^{s_1^0} d s_1
 \int^{s_2^0} d s_2\frac{\rho^{pert}_i(s_1,s_2,q^2)}{(s_1-p_1^2)(s_2-p_2^2)},
 \nonumber  \\
f_i^{(4)}&=&\int^{s_1^0} d s_1
 \int^{s_2^0}d s_2\frac{\rho^{(4)}_i(s_1,s_2,q^2)}{(s_1-p_1^2)(s_2-p_2^2)}.
 \nonumber  \nonumber
 \end{eqnarray}

\subsection*{2. The Calculation of the Wilson Coefficients}

In this section, we calculate the Wilson coefficients in the
operator-product expansion, then extract the relevant
coefficients $f_i$ for the sum rules of the form factors in
Eq.~(\ref{21}). The method of the calculation is very similar
to that used in our previous work in Ref. \cite{R13}, where
the form factors in $D_s^+\to \phi\bar{\ell}\nu$ decay were studied in QCD sum rule
method. So we will not give the details of the calculation in the present paper. But
for the completeness of this paper we shall give some main points of the calculation in this section.

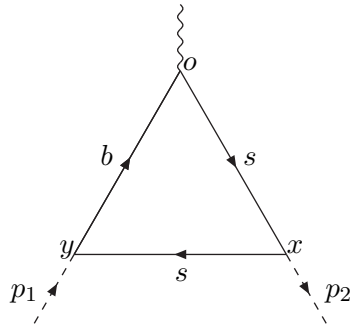
\begin{figure}[h]
  \begin{center}
    \begin{picture}(100,80)(-15,-25.98)
    \DashArrowLine(-15,-25.98)(0,0){3}\ArrowLine(0,0)(40,69.28)
    \ArrowLine(0,0)(40,69.28)\ArrowLine(40,69.28)(80,0)\ArrowLine(80,0)(0,0)
    \DashArrowLine(80,0)(95,-25.98){3}\Photon(40,69.28)(40,94.28){1}{4}
    \put(41,70){$o$}
    \put(80.5,0.5){$x$}
    \put(-5.5,0.5){$y$}
    \put(-24,-16){$p_1$}
    \put(95,-16){$p_2$}
    \put(10,34){$b$}
    \put(38,-10){$s$}
    \put(64,34){$s$}
    \end{picture}
    \caption{\label{fig:pert}\small Diagram for perturbative contribution.}
  \end{center}
  \end{figure}

\begin{figure}[h]
\begin{center}
\begin{picture}(300,80)(-15,-25.98)
\DashArrowLine(-15,-25.98)(0,0){3} \ArrowLine(0,0)(40,69.28)
\ArrowLine(0,0)(40,69.28)\ArrowLine(40,69.28)(80,0)\ArrowLine(80,0)(50,0)\ArrowLine(30,0)(0,0)
\DashArrowLine(80,0)(95,-25.98){3}\Photon(40,69.28)(40,94.28){1}{4}\put(47,-2.7){$\times$}
\put(26.8,-2.7){$\times$}\put(41,70){$o$}\put(80.5,0.5){$x$}\put(-5.5,0.5){$y$}

\DashArrowLine(185,-25.98)(200,0){3} \ArrowLine(200,0)(240,69.28)
\ArrowLine(200,0)(240,69.28)\ArrowLine(240,69.28)(255,43.3)
\ArrowLine(265,25.98)(280,0) \ArrowLine(280,0)(200,0)
\DashArrowLine(280,0)(295,-25.98){3}\Photon(240,69.28)(240,94.28){1}{4}
\put(250.2,41.6){$\times$}\put(261,23.8){$\times$}
\put(241,70){$o$}\put(280.5,0.5){$x$}\put(194.5,0.5){$y$}
\put(33,-40){$(a)$}\put(233,-40){$(b)$}
\end{picture}
\end{center}
\vspace{0.5cm}
\caption{\label{biquark}\small Diagrams for the contributions of operators $\bar{\Psi}(x)\Psi(y)$ and $\bar{\Psi}(0)\Psi(x)$.}
\end{figure}
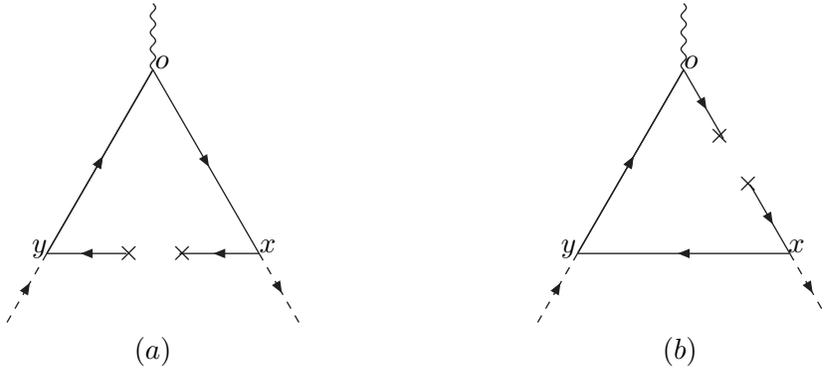

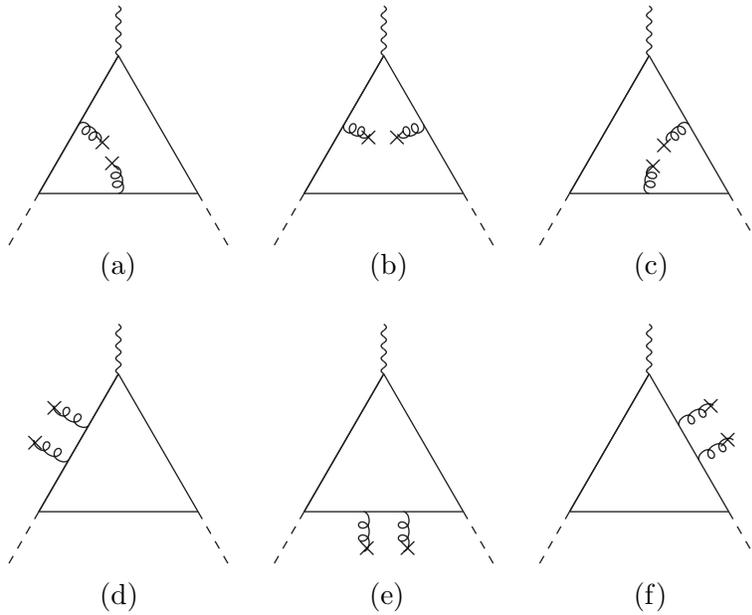
\begin{figure}[h]
\begin{center}
\begin{picture}(300,200)(-15,-25.98)
\DashLine(-11.25,100.515)(0,120){3} \Line(0,120)(30,171.96)
\Line(0,120)(30,171.96)\Line(30,171.96)(60,120)\Line(60,120)(0,120)
\DashLine(60,120)(71.25,100.515){3}\Photon(30,171.96)(30,190.71){1}{4}

\GlueArc(0,120)(30,40,60){2}{2}\GlueArc(0,120)(30,0,20){2}{2}
\put(19.88,137,25){$\times$}\put(23.5,129){$\times$}

\DashLine(88.75,100.515)(100,120){3} \Line(100,120)(130,171.96)
\Line(100,120)(130,171.96)\Line(130,171.96)(160,120)\Line(160,120)(100,120)
\DashLine(160,120)(171.25,100.515){3}\Photon(130,171.96)(130,190.71){1}{4}

\GlueArc(130,171.96)(30,240,260){2}{2}\GlueArc(130,171.96)(30,280,300){2}{2}
\put(120.4,138.8){$\times$}\put(131.3,138.8){$\times$}

\DashLine(188.75,100.515)(200,120){3} \Line(200,120)(230,171.96)
\Line(200,120)(230,171.96)\Line(230,171.96)(260,120)\Line(260,120)(200,120)
\DashLine(260,120)(271.25,100.515){3}\Photon(230,171.96)(230,190.71){1}{4}

\GlueArc(260,120)(30,160,180){2}{2}\GlueArc(260,120)(30,120,140){2}{2}
\put(232.3,136){$\times$}\put(228.125,128){$\times$}

\DashLine(-11.25,-19.485)(0,0){3} \Line(0,0)(30,51.96)
\Line(0,0)(30,51.96)\Line(30,51.96)(60,0)\Line(60,0)(0,0)
\DashLine(60,0)(71.25,-19.485){3}\Photon(30,51.96)(30,70.71){1}{4}

\Gluon(11.25,19.485)(-1.74,26.985){2}{2}\Gluon(18.75,32.475)(5.76,39.975){2}{2}

\put(-5.625,23.5){$\times$}\put(1.65,36.5){$\times$}
 \DashLine(88.75,-19.485)(100,0){3} \Line(100,0)(130,51.96)
\Line(100,0)(130,51.96)\Line(130,51.96)(160,0)\Line(160,0)(100,0)
\DashLine(160,0)(171.25,-19.485){3}\Photon(130,51.96)(130,70.71){1}{4}

\Gluon(122.5,0)(122.5,-15){2}{2}\Gluon(137.5,0)(137.5,-15){2}{2}

\put(119.875,-16.5){$\times$}\put(135.25,-16.5){$\times$}

\DashLine(188.75,-19.485)(200,0){3} \Line(200,0)(230,51.96)
\Line(200,0)(230,51.96)\Line(230,51.96)(260,0)\Line(260,0)(200,0)
\DashLine(260,0)(271.25,-19.485){3}\Photon(230,51.96)(230,70.71){1}{4}

\Gluon(248.75,19.485)(261.74,26.985){2}{2}\Gluon(241.25,32.475)(254.24,39.975){2}{2}

\put(256.5,24.75){$\times$}\put(250,37.5){$\times$}

\put(23,90){(a)}
 \put(125,90){(b)}
 \put(225,90){(c)}

 \put(23,-35){(d)}
 \put(125,-35){(e)}
 \put(225,-35){(f)}
\end{picture}
\end{center}
\vspace{0.5cm}
\caption{\label{gg}\small Diagrams for contributions of gluon-gluon operator.}
\end{figure}

\begin{figure}[h]
\begin{center}
\begin{picture}(400,80)(-15,-25.98)
\DashLine(-11.25,-19.485)(0,0){3} \Line(0,0)(30,51.96)
\Line(0,0)(30,51.96)\Line(30,51.96)(60,0)\Line(60,0)(37.5,0)\Line(22.5,0)(0,0)
\DashLine(60,0)(71.25,-19.485){3}\Photon(30,51.96)(30,70.71){1}{4}\put(35.25,-2.025){$\times$}
\put(19,-2.025){$\times$}
\GlueArc(0,0)(30,20,60){2}{3}\put(25.5,10){$\times$}

\DashLine(88.75,-19.485)(100,0){3} \Line(100,0)(130,51.96)
\Line(100,0)(130,51.96)\Line(130,51.96)(160,0)\Line(160,0)(137.5,0)\Line(122.5,0)(100,0)
\DashLine(160,0)(171.25,-19.485){3}\Photon(130,51.96)(130,70.71){1}{4}\put(135,-2.025){$\times$}
\put(119,-2.025){$\times$}
\GlueArc(160,0)(30,120,160){2}{3}\put(127.5,8.5){$\times$}
\DashLine(188.75,-19.485)(200,0){3} \Line(200,0)(230,51.96)
\Line(200,0)(230,51.96)\Line(230,51.96)(241.25,32.475)
\Line(248.75,19.485)(260,0) \Line(260,0)(200,0)
\DashLine(260,0)(271.25,-19.485){3}\Photon(230,51.96)(230,70.71){1}{4}
\put(236.5,31.2){$\times$}\put(245,17){$\times$}

\GlueArc(260,0)(30,140,180){2}{3}\put(233.5,17.5){$\times$}
\DashLine(288.75,-19.485)(300,0){3} \Line(300,0)(330,51.96)
\Line(300,0)(330,51.96)\Line(330,51.96)(341.25,32.475)
\Line(348.75,19.485)(360,0) \Line(360,0)(300,0)
\DashLine(360,0)(371.25,-19.485){3}\Photon(330,51.96)(330,70.71){1}{4}

\put(336.5,31.2){$\times$}\put(345,17){$\times$}

\GlueArc(330,51.96)(30,240,280){2}{3}\put(332.25,19.5){$\times$}
\put(22,-28){(a)}
 \put(122,-28){(b)}
 \put(222,-28){(c)}

\put(322,-28){(d)}
\end{picture}
\end{center}
\caption{\label{qg}Diagrams for mixed  quark-gluon operators.}
\end{figure}
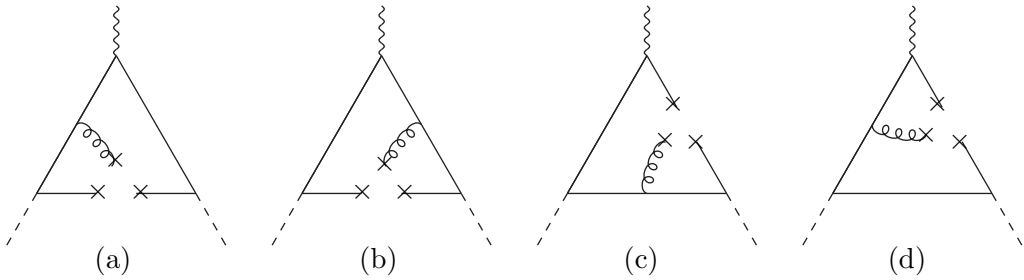

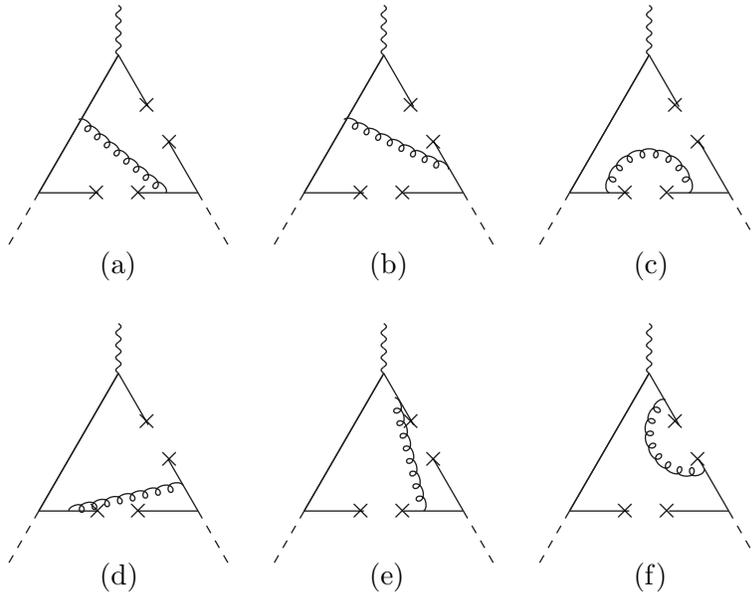
\begin{figure}[h]
\begin{center}
\begin{picture}(300,220)(-15,-25.98)
\DashLine(-11.25,100.515)(0,120){3} \Line(0,120)(30,171.96)
\Line(0,120)(30,171.96)\Line(60,120)(37.5,120)\Line(22.5,120)(0,120)
\DashLine(60,120)(71.25,100.515){3}\Photon(30,171.96)(30,190.71){1}{4}
\put(33.3,117.975){$\times$}
\put(17.5,117.975){$\times$} \Line(30,171.96)(41.25,152.475)
\Line(48.75,139.485)(60,120)
\put(36.5,151.2){$\times$}\put(45,137){$\times$}

\Gluon(15,147.6)(48,120){1.6}{8}
 \DashLine(88.75,100.515)(100,120){3} \Line(100,120)(130,171.96)
\Line(100,120)(130,171.96)\Line(160,120)(137.5,120)\Line(122.5,120)(100,120)
\DashLine(160,120)(171.25,100.515){3}\Photon(130,171.96)(130,190.71){1}{4}
\put(133.3,117.975){$\times$} \put(117.5,117.975){$\times$}
\Line(130,171.96)(141.25,152.475) \Line(148.75,139.485)(160,120)
\put(136.5,151.2){$\times$}\put(145,137){$\times$}

\Gluon(115,147.6)(154.375,128.8){1.6}{8}
\DashLine(188.75,100.515)(200,120){3} \Line(200,120)(230,171.96)
\Line(200,120)(230,171.96)\Line(260,120)(237.5,120)\Line(222.5,120)(200,120)
\DashLine(260,120)(271.25,100.515){3}\Photon(230,171.96)(230,190.71){1}{4}
\put(233.3,117.975){$\times$}
\put(217.5,117.975){$\times$} \Line(230,171.96)(241.25,152.475)
\Line(248.75,139.485)(260,120)
\put(236.5,151.2){$\times$}\put(245,137){$\times$}

\GlueArc(230,120)(15,0,180){1.6}{8} \DashLine(-11.25,-19.485)(0,0){3}
\Line(0,0)(30,51.96)
\Line(0,0)(30,51.96)\Line(60,0)(37.5,0)\Line(22.5,0)(0,0)
\DashLine(60,0)(71.25,-19.485){3}\Photon(30,51.96)(30,70.71){1}{4}
\put(33.3,-2.5){$\times$}
\put(18,-2.5){$\times$} \Line(30,51.96)(41.25,32.475)
\Line(48.75,19.485)(60,0)
\put(36.5,31.2){$\times$}\put(45,17){$\times$}
\Gluon(11.25,0)(54.375,9.675){1.6}{8}
\DashLine(88.75,-19.485)(100,0){3} \Line(100,0)(130,51.96)
\Line(100,0)(130,51.96)\Line(160,0)(137.5,0)\Line(122.5,0)(100,0)
\DashLine(160,0)(171.25,-19.485){3}\Photon(130,51.96)(130,70.71){1}{4}
\put(133.3,-2.5){$\times$}
\put(117.5,-2.5){$\times$} \Line(130,51.96)(141.25,32.475)
\Line(148.75,19.485)(160,0)
\put(136.5,31.2){$\times$}\put(145,17){$\times$}

\Gluon(134.125,42.75)(145,0){1.6}{8}
\DashLine(188.75,-19.485)(200,0){3} \Line(200,0)(230,51.96)
\Line(200,0)(230,51.96)\Line(260,0)(237.5,0)\Line(222.5,0)(200,0)
\DashLine(260,0)(271.25,-19.485){3}\Photon(230,51.96)(230,70.71){1}{4}
\put(233.3,-2.5){$\times$}
\put(217.5,-2.5){$\times$} \Line(230,51.96)(241.25,32.475)
\Line(248.75,19.485)(260,0)
\put(236.5,31.2){$\times$}\put(245,17){$\times$}

\GlueArc(245,29.7)(15,126,293){1.6}{8}

\put(23,90){(a)}
 \put(125,90){(b)}
 \put(225,90){(c)}

 \put(23,-28){(d)}
 \put(125,-28){(e)}
 \put(225,-28){(f)}
\end{picture}
\end{center}
\caption{\label{4q}\small Diagrams for four-quark contributions.}
\end{figure}

All of the Feynman diagrams for calculating the Wilson coefficients in OPE in Eq. (\ref{17}) are shown below. They are: diagram for perturbative contribution in Fig.\ref{fig:pert}, diagrams for contributions of operators $\bar{\Psi}(x)\Psi(y)$ and $\bar{\Psi}(0)\Psi(x)$  in Fig.\ref{biquark},
diagrams for contributions of gluon-gluon operator in Fig.\ref{gg}, diagrams for mixed  quark-gluon operators in Fig.\ref{qg} and diagrams for contributions of four-quark operators in Fig.\ref{4q}.

\newpage

\subsubsection*{2.1 The perturbation contribution}

For perturbative contribution shown in
Fig.\ref{fig:pert},  only the leading order in $\alpha_s$ expansion is considered
here. This contribution is relevant to the Wilson coefficient $C_0$ in OPE
of the correlation function in Eq.~(\ref{18}). The amplitude can be written as
\begin{eqnarray}
C_0&=&i^2\int
\frac{d^4k}{(2\pi)^4}(-1)Tr\left[i\gamma_5\frac{i(\not{k}+m_s)}{k^2-m_s^2+i\varepsilon}
\gamma_\mu
\frac{i(\not{k}+\not{p}_2+m_s)}{(k+p_2)^2-m_s^2+i\varepsilon}\gamma_\nu
(1-\gamma_5)\right.\nonumber \\
&&\left.\frac{i(\not{k}+\not{p}_1+m_s)}{(k+p_1)^2-m_b^2+i\varepsilon}\right].
\label{22}
\end{eqnarray}

We can re-write the integration of Eq.~(\ref{22}) in the form of dispersion integration
\begin{equation}\label{23}
C_0=\int d s_1 d s_2
\frac{\rho(s_1,s_2,q^2)}{(s_1-p_1^2)(s_2-p_2^2)}.
\end{equation}
The spectral density $\rho(s_1,s_2,q^2)$ can be calculated according to Cutkosky's rule \cite{Cutkosky}, i.e., replacing the denominators of the quark
propagators with $\delta$ functions and putting all the quark lines on-mass-shell,
$1/(k^2-m^2+i\varepsilon)\to -2\pi i \delta(k^2-m^2)$.
Then the spectral density can be calculated from
\begin{eqnarray}\label{25}
\rho(s_1,s_2,q^2)&=&\frac{(-2\pi i)^3}{-4{\pi}^2}
\int\frac{d^4k}{(2\pi)^4} Tr[\gamma_5(\not{k}+m_s)\gamma_\mu
(\not{k}+\not{p}_2+m_s)\gamma_\nu (1-\gamma_5)\nonumber \\
&&\times
(\not{k}+\not{p}_1+m_b)]\delta (k^2-m_0^2)\delta[(k+p_1)^2-m_1^2]\nonumber \\
&& \times\delta [(k+p_2)^2-m_2^2]_{\left|\small{\begin{array}{ll}
                                p_1^2\to s_1, p_2^2\to s_2
                                \end{array}}\right.} .
\end{eqnarray}
Some basic formulas are needed to perform the integration in Eq.~(\ref{25}),
which have been obtained in Refs.~\cite{R6,R13}. They are given in Appendix A. With results of $I$, $I_\mu$ and $I_{\mu\nu}$ given in
Eqs.~(A1) $\sim$ (A3), the integration in Eq.~(\ref{25}) can be performed without difficulty.

\subsubsection*{2.2   Contributions of the non-local quark-quark operator}

The diagrams for the contribution of non-local ``quark-quark" operator are shown in Fig.\ref{biquark}. The contribution of Fig.\ref{biquark} (b) is zero after the double Borel transformation for both
variables $p_1^2$ and $p_2^2$, because only one variable left in the denominator $1/(p_2^2-m_s^2)$. So Fig.\ref{biquark} (b) can be ignored.

The contribution of Fig.\ref{biquark}(a) is
\begin{equation}\label{29}
\Pi_{\mu\nu}^{2a}=i^2\int d^4 x d^4 y e^{ip_2\cdot x-ip_1\cdot y}
  \langle 0 |\bar{\Psi}(x)\gamma_\mu S_{F}^s(x)\gamma_\nu
  (1-\gamma_5) S_F^b(-y)i\gamma_5 \Psi (y)|0\rangle ,
\end{equation}
where
$$S_{F}^s(x)=\int\frac{d^4k_2}{(2\pi)^4}\frac{i}{\not{k_2}-m_s}e^{-ik_2x},
~S_{F}^b(-y)=\int\frac{d^4k_1}{(2\pi)^4}\frac{i}{\not{k_1}-m_b}e^{ik_1y}$$
are the propagators of $s$
and $b$ quarks, respectively.  Moving the quark field operators
$\bar{\Psi}(x)$ and $\Psi(y)$ together then we can obtain
\begin{equation}
\Pi_{\mu\nu}^{2a}=i^2\int d^4 x d^4 y e^{ip_2\cdot x-ip_1\cdot y}
  \langle 0 |\bar{\Psi}_\alpha (x) \Psi_\beta (y)  |0\rangle
  [\gamma_\mu S_{F}^s(x)\gamma_\nu (1-\gamma_5) S_F^b(-y)i\gamma_5]_{\alpha\beta} ,
  \label{30}
\end{equation}
where $\alpha$ and $\beta$ are Dirac spinor indices. The matrix
element $\langle 0 |\bar{\Psi}_\beta (x) \Psi_\alpha (y)
|0\rangle$ can be treated in the fixed-point gauge \cite{Schwinger1973,Shif1980,Dub1981},
the result of which up to the order of $x^3$ and $y^3$ has been given in Ref.~\cite{R13}
\begin{eqnarray}
&&\langle 0 |\bar{\Psi}^a_\alpha (x)\Psi^b_\beta (y)|0\rangle
 =\delta_{ab}\left[ \langle \bar{\Psi} \Psi\rangle\left(
   \frac{1}{12}\delta_{\beta\alpha}+i\frac{m}{48}
   (\not{x}-\not{y})_{\beta\alpha}-\frac{m^2}{96}
 (x-y)^2\delta_{\beta\alpha}\right.\right. \nonumber\\
   &&\left.\left.-\frac{i}{3!}\frac{m^3}{96}
 (x-y)^2(\not{x}-\not{y})_{\beta\alpha} \right)
 +g\langle \bar{\Psi}\sigma TG\Psi\rangle\left(\frac{1}{192}(x-y)^2
 \delta_{\beta\alpha}\right.\right. \nonumber\\
   &&\left.\left.+\frac{i}{3!}\frac{m}{192}
 (x-y)^2(\not{x}-\not{y})_{\beta\alpha}\right)
 -\frac{i}{3!}\frac{g^2}{3^4\times 2^4}\langle
 \bar{\Psi}\Psi\rangle ^2 (x-y)^2(\not{x}-\not{y})_{\beta\alpha}
 \right. \nonumber \\&& \left.\frac{}{}+\cdots \right],
 \label{31}
 \end{eqnarray}
where $a$ and $b$ in the above equation are the color indices, $m$ is the quark mass.
From Eq.~(\ref{31}) one can see that Fig.\ref{biquark} (a)
contributes not only to the coefficients of quark condensate $\langle
\bar{\Psi}\Psi\rangle$, but also to mixed quark-gluon condensate $g\langle
\bar{\Psi}\sigma TG\Psi\rangle$ and the four-quark condensate
$\langle \bar{\Psi}\Psi\rangle ^2$.


 \subsubsection*{2.3   Contributions of the non-local gluon-gluon operator}

The diagrams for the contribution of non-local gluon-gluon operator are
shown in Fig.\ref{gg}. It is convenient to calculate these diagrams in the fixed-point gauge,
in which the gauge fixing condition is taken as
$z^\mu A^a_\mu(z)=0$ \cite{Schwinger1973,Shif1980,Dub1981}. Then the external color field
can be expressed in terms of its strength tensor \cite{Shif1980},
 \begin{equation}\label{32}
A^a_\mu (z)=\int^1_0 d\beta \beta z^\rho G^a_{\rho\mu}(\beta
z).
\end{equation}
Expanding the above expression to the first order of $z$, one can get
\begin{equation}\label{33}
A^a_\mu (z)=\frac{1}{2}z^\rho  G^a_{\rho\mu}(0)+\cdots.
\end{equation}

Another equation useful for the calculation of the contributions of the diagrams
depicted in Fig.\ref{gg} is
\begin{equation}\label{34}
\langle 0|G^a_{\alpha\sigma}G^b_{\beta\rho}|0\rangle=\frac{1}{96}
 \langle GG \rangle \delta_{ab}(g_{\alpha\beta}g_{\sigma\rho}-
  g_{\alpha\rho}g_{\sigma\beta}),
\end{equation}
where $\langle GG \rangle$ is the abbreviation of
$\langle 0|G^a_{\mu\nu}G^{a\mu\nu}|0\rangle$. Using this equation we can decompose
the matrix element $\langle 0|G^a_{\alpha\sigma}G^b_{\beta\rho}|0\rangle$ to obtain the gluon-gluon
condensate.

It has been shown that the sum of the contributions of the diagrams in Fig.\ref{gg} cancel
in $D_S\to\phi$ transitions in Ref.~\cite{R13}. Similar case occurs in the calculation for
$\bar{B}^0_s\to\phi$ transitions. Therefore there are still no contributions of gluon-gluon condensate in
$\bar{B}^0_s\to\phi$ transition.

 \subsubsection*{2.4   Contributions of the non-local quark-gluon mixing and four-quark operators}

The diagrams for the contribution of non-local quark-gluon mixing and four-quark operators
are shown in Figs.\ref{qg} and \ref{4q}, respectively. The methods to calculate the
contributions of these diagrams are similar to that for other diagrams. Two different
vacuum-expectation values of the non-local quark-gluon mixing operators should be used. They are:

(1) The vacuum expectation value of quark-gluon mixing operator
$\bar{\Psi}(x) \Psi (y)G^a_{\mu\nu}$, which is expanded to be \cite{R13}
\begin{eqnarray}\label{35}
&&\langle 0|\bar{\Psi}^i_\alpha (x) \Psi ^j_\beta (y)
G^a_{\mu\nu}|0\rangle \nonumber\\
&=&\frac{1}{192}\langle \bar{\Psi}\sigma T G\Psi \rangle
(\sigma_{\mu\nu})_{\beta\alpha}T^a_{ji} + \left[-\frac{g}{96\times
9}\langle\bar{\Psi}\Psi\rangle^2
(g_{\rho\mu}\gamma_\nu-g_{\rho\nu}\gamma_\mu)(x+y)^\rho\right.\nonumber\\
&&\left.+i(y-x)^\rho\left(\frac{g}{96\times
9}\langle\bar{\Psi}\Psi\rangle^2 +\frac{m}{96\times 4}\langle
\bar{\Psi}\sigma T G\Psi \rangle
\right)\varepsilon_{\rho\mu\nu\sigma}\gamma_5\gamma^\sigma
\right]_{\beta\alpha}T^a_{ji},
\end{eqnarray}
where $\langle \bar{\Psi}\sigma TG\Psi \rangle$ and
$\langle\bar{\Psi}\Psi\rangle^2$ are the abbreviations of $\langle
0| \bar{\Psi}\sigma_{\mu\nu} T^aG^{a\mu\nu}\Psi |0 \rangle$ and
$\langle 0|\bar{\Psi}\Psi |0\rangle^2$ respectively, and $g$ is the
strong coupling constant.

(2) The other matrix element needed is \cite{R6}
\begin{equation}\label{37}
\langle 0|\bar{\Psi}^i_\alpha \Psi^j_\beta \hat{D}_\xi
G^a_{\sigma\rho} |0\rangle=-\frac{g}{3^3\times 2^4}
\langle \bar{\Psi}\Psi \rangle ^2
(g_{\xi\rho}\gamma_\sigma-g_{\xi\sigma}\gamma_\rho )_{\beta\alpha}
T^a_{ji}~,
\end{equation}
and the external color field in fix-point gauge expanded up to the second order is used
\begin{eqnarray}\label{36}
A^a_{\mu}(z)&=&\int^1_0d\beta \beta z^\rho
G^a_{\rho\mu}(\beta z)\nonumber\\
&=&\frac{1}{2}z^\rho G^a_{\rho\mu}(0)+\frac{1}{3}z^\alpha z^\rho
\hat{D}_\alpha G^a_{\rho\mu}(0)+\cdots ~,
\end{eqnarray}
here $\hat{D}_\alpha$ is the covariant derivative in the adjoint
representation, $(\hat{D}_\alpha)^{mn} =\partial_\alpha
\delta^{mn}-gf^{amn}A^a_\alpha$.

After calculating all of the diagrams in  Figs.\ref{qg} and \ref{4q}, we find that the contributions of
Fig.\ref{qg} (c), (d) and Fig.\ref{4q} (c), (d) are vanishes after double Borel
transformation for both variables $p_1^2$ and $p_2^2$, because only
one variable appearing in the denominator. For example,
$1/{q^2 (p_1^2-m_1^2)}$. The Borel transformation for $p_2^2$
will eliminate such terms.

Using the above method, we get the coefficients $\hat{B}f_0$, $\hat{B}(f_1+f_3)$, $\hat{B}(f_1-f_3)$ and $\hat{B}f_5$ needed in Eq.~(\ref{21}), which are given in the Appendix B.

\section{Numerical Calculation of the Form Factors}

For the numerical calculation, the standard values of the
condensates at the renormalization point $\mu =1\mbox{GeV}$ are
used \cite{SVZ1,SVZ2,Colan2000},
\begin{eqnarray}\label{38}
&\langle \bar{q}q\rangle =-(0.24\pm 0.01 \mbox{GeV})^3, ~~~~
\langle \bar{s}s\rangle =(0.8\pm 0.2)\langle \bar{q}q\rangle,
\nonumber\\[4mm]
 & g\langle \bar{\Psi}\sigma TG\Psi \rangle =m_0^2 \langle
\bar{\Psi}\Psi \rangle, ~~~~\alpha_s\langle \bar{\Psi}\Psi\rangle
^2= 6.0\times10^{-5}\mbox{GeV}^6 ,\\[4mm]
& m_0^2=0.8\pm 0.2 \mbox{GeV}^2. \nonumber
\end{eqnarray}
The quark masses are $m_s=95~\mbox{MeV}$,
$m_b=4.18~\mbox{GeV}$ \cite{R17}, the meson masses are $m_\phi=1.02~\mbox{GeV}$,
$m_{J/\psi}=3.097~\mbox{GeV}$, $m_{B_s}=5.367~\mbox{GeV}$ \cite{R17}. The decay constants for $\phi$ and $J/\psi$ mesons are extracted from the experimental measurement of the branching ratios of $\phi \to \ell^+\ell$ and $J/\psi \to \ell^+\ell$ \cite{R17}, which are $f_{\phi}=0.228~\mbox{GeV}$ and $f_{J/\psi}=0.416~\mbox{GeV}$.
For the decay constant of $B_s$ meson we take $f_{B_s}=0.266\pm 0.019\mbox{GeV}$ \cite{R16,fBs}. The threshold parameters $s_1^0$ and $s_2^0$ for $B_s$ and $\phi$ mesons are taken to be $s_1^0=34.9-35.9\mbox{GeV}^2$, $s_2^0=1.9-2.1
\mbox{GeV}^2$, respectively.

The physical result should not depend on the Borel parameters $M_1$ and $M_2$ if the OPE
were calculated up to infinite order. However, in practice OPE can only be calculated up to finite orders.
So Borel parameters have to be selected in some ``windows"
to get the best stability of the physical results.
The criterion to choose the region for $M_1$ and $M_2$ is:
(1) The contributions of the excited and continuum states should be effectively
suppressed to make sure that the sum rule does not depend on the approximation for the
excited and continuum states sensitively. This requires that the Borel parameters
should not be too large; (2) The contribution of the condensates of
higher dimensional operators should be small to make sure the truncated OPE is effective.
The series in OPE generally depends on Borel parameters in the denominator $1/M^n_{1,2}$, where $n$ is
positive integer. The higher the dimension of the operator, the larger the integer $n$.
This requires that the Borel parameters should not be too small.

\begin{figure}[tbp]
  \centering
  \includegraphics[width=0.85\textwidth,origin=l,angle=0]{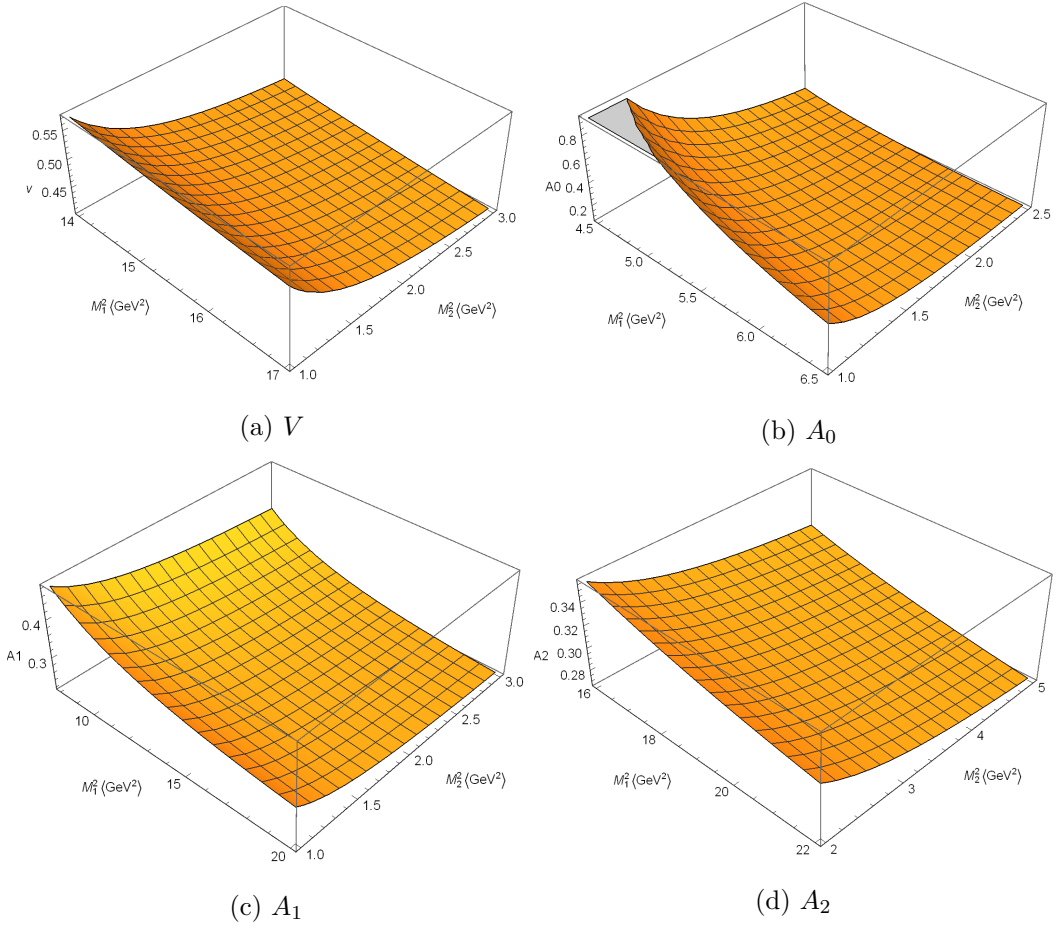}
  \caption{Form factors changing with $M_1^2$ and $M_2^2$.} \label{3-dim}
\end{figure}

\begin{figure}[tbp]
  \centering
  \includegraphics[width=0.85\textwidth,origin=l,angle=0]{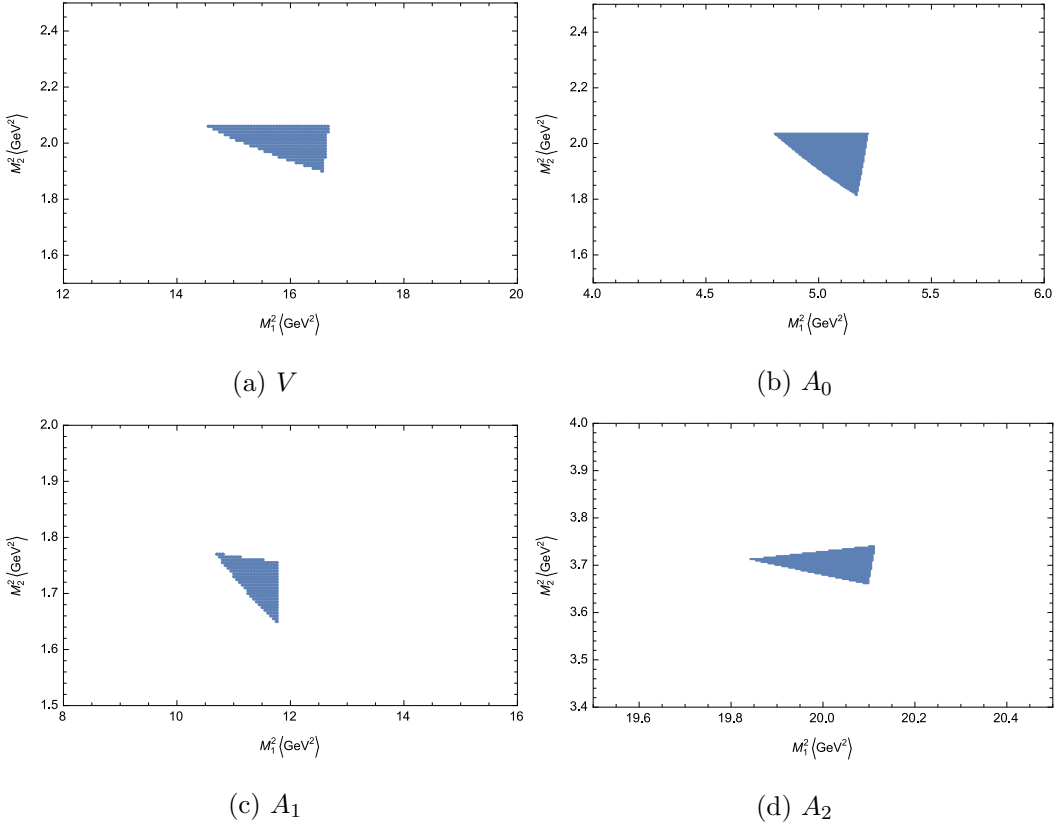}
  \caption{\label{window} Selected regions of $M_1^2$ and $M_2^2$.}
\end{figure}

After numerical analysis, we find the optimal stability in accord with the requirements shown in Table \ref{T1}. The three-dimensional diagrams of form factors changing with $M_1^2$ and $M_2^2$ are depicted in Fig.\ref{3-dim}. The stability regions relevant to the requirements in Table \ref{T1} are shown in Fig.\ref{window} as two-dimensional diagram of $M_1^2$ and $M_2^2$. Combining
Fig.\ref{3-dim} and Fig.\ref{window}, we can find good stabilities for the form factors
within these regions.
\begin{tiny}
\begin{table}[h]
\caption{Requirements to select Borel Parameters $M_1^2$ and $M_2^2$
 for each form factors $V(0)$, $A_0(0)$, $A_1(0)$ and $A_2(0)$}
\begin{center}
\begin{tabular}{|c|c|c|c|}\hline
Form Factors & contribution  & continuum of & continuum of  \\
             &of condensate &  $B_s$ channel & $\phi$ channel\\
             \hline
$V(0)$  & $\le 56.7\% $ & $\le 11\%$ & $\le 56\%$ \\ \hline $A_0(0)$&
$\le 14\% $ & $\le 10\%$ & $\le 50\%$ \\ \hline $A_1(0)$ &$\le
56\% $ & $\le 17.5\%$ & $\le 50\%$ \\ \hline $A_2(0)$  & $\le 5.2\% $
& $\le 17.2\%$ & $\le 54\%$ \\ \hline
\end{tabular}\end{center}
\label{T1}
\end{table}
\end{tiny}

The final results for the form factors at $q^2=0$ are
\begin{eqnarray}\label{39}
V(0)&=& 0.45\pm 0.10,~~~~~~~~A_0(0)~=~0.30\pm 0.25,\nonumber
\\[4mm]
A_1(0)&=& 0.32\pm 0.07,~~~~~~~A_2(0)~=~0.30\pm 0.07.
\end{eqnarray}
The uncertainties are obtained by varying the input parameters and Borel parameters in the stability regions.

We compare our results with other nonperturbative approaches such as LCSR \cite{R11} and CQM \cite{R12} in Table \ref{T2}. The form factors for semileptonic decays of $B^0_s$ to $\phi$ meson have also been calculated by QCD sum rule in Ref. \cite{R15}. We do not list the value of $A^\star_2(0)=-0.44$ of Ref. \cite{R15} in Table \ref{T2}, because the form factor $A^\star_2(0)$ defined in \cite{R15} does not directly correspond to the defination in our work. The relations of the form factors defined in Ref.\cite{R15} and ours are
\begin{eqnarray}\label{relation}
&V^\star(q^2)=(-i)V(q^2),~~{A^\star_0}(q^2)=(-i){A_1}(q^2),~~{A^\star_1}(q^2)=(-i){A_2}(q^2),\nonumber\\
&{A^\star_2}(q^2)=\frac{(-i)(m_{B_s}+m_\phi)}{q^2}\left[(m_{B_s}-m_\phi)A_2(q^2)-(m_{B_s}+m_\phi)A_1(q^2)+2m_\phi A_1(q^2)\right],
\end{eqnarray}
where $V^\star(q^2)$, ${A^\star_0}(q^2)$, ${A^\star_1}(q^2)$ and ${A^\star_2}(q^2)$ denote the form factors defined in \cite{R15}.

\begin{tiny}
\begin{table}[h]
\caption{Comparison of our results of form factors with other work}
\begin{center}
\begin{tabular}{|c|c|c|c|c|}
\hline
 &$A_0(0)$ & $A_1(0)$ &$A_2(0)$ &$V(0)$ \\
                         \hline
$\mbox{LCSR}$  & $0.474 $ & $0.311$ & $0.234$ & $0.434$
\\ \hline
$\mbox{CQM}$  & $0.42 $&$0.34$ & $0.31$ & $0.44$ \\ \hline
$\mbox{SR}$  & $A^\star_2(0)$ & $-0.34$ & $0.35$ & $-0.47$ \\ \hline
$\mbox{This~work}$    & $0.30\pm0.25 $ & $0.32\pm0.07$ & $0.30\pm0.07$ & $0.45\pm0.10$ \\ \hline
\end{tabular}\end{center}
\label{T2}
\end{table}
\end{tiny}

Table \ref{T2} shows that our results for $A_1$, $A_2$ and $V$ are more consistent with the results of LCSR in Ref. \cite{R11} and  CQM in Ref. \cite{R12}. Only $A_0$ is slightly smaller than theirs. The difference between the results of the form factors in Ref. \cite{R15} and ours is large. The reason is checked, that is: for the contribution of the condensate of the operator of dimension 3, the leading contribution is at the order of $(m_s/M_i)^0$ in our calculation, which comes from the first term $\frac{1}{12}\delta_{\beta\alpha}$ of Eq. (\ref{31}). But there are no such terms in the result of Ref. \cite{R15}, only terms like $(m_b m_s/M_i^2)^n$ or $(m_s^2/M_i^2)^n$ with $n\ge 1$ exist. The contributions of the operators of dimension 5 are also different.

In the next section we can see that the branching ratios of $B_s\to J\psi \phi$ calculated with the form factors obtained in this work are consistent with experimental data.

For the $q^2$-dependence of the form factors, we varied the value of $q^2$ by keeping it slightly larger than 0. We find that the $q^2$-dependence of $V(q^2)$, $A_0(q^2)$ and $A_2(q^2)$ are well compatible with the pole-model \cite{R1}, which can be expressed as
\begin{eqnarray}
V(q^2)~=&\frac{V(0)}{1-q^2/(m_{pole}^V)^2},  \nonumber\\
A_0(q^2)~=&\frac{A_0(0)}{1-q^2/(m_{pole}^{A_0})^2}, \nonumber\\
A_2(q^2)~=&\frac{A_2(0)}{1-q^2/(m_{pole}^{A_2})^2},
\end{eqnarray}
while the $q^2$ dependence of $A_1(q^2)$ is very weak.

We fit $V(q^2)$, $A_0(q^2)$ and $A_2(q^2)$ with the pole model to our numerical results calculated from
QCD sum rule. Then the relevant fitted pole masses are
\begin{eqnarray}\label{40}
m^V_{pole}~=& 5.59\pm 0.27~\mbox{GeV},\nonumber\\
m^{A_0}_{pole}~=& 5.62\pm 2.38~\mbox{GeV},\nonumber\\
m^{A_2}_{pole}~=& 9.20\pm 0.40~\mbox{GeV}.
\end{eqnarray}
The fitted pole mass $m^{A_2}_{pole}$ is apparently larger than the other two pole masses,
which means that the dependence of $A_2(q^2)$ on $q^2$ is also weak, and it is almost as weak as $A_1(q^2)$.
This result implies that the dependence of the form factors
on $q^2$ can not always be described by a real physical resonance pole that is associated with the transition
current, because the mass of such resonance is usually far beyond the physical
region of $q^2$ in the realistic decay process.

\section{The Application of the Form Factors to the Branching Ratios}
\begin{figure}[tbp]
  \centering
  \includegraphics[width=0.60\textwidth,origin=l,angle=0]{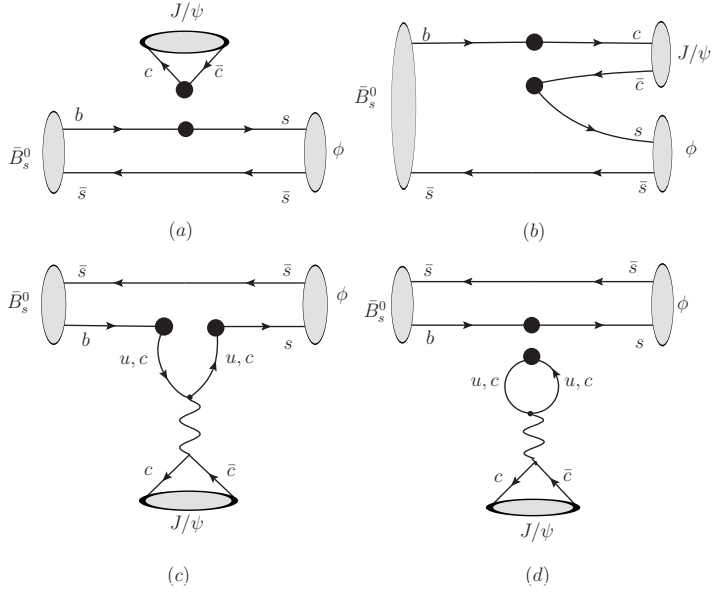}
  \caption{The Feynman diagrams for the decay of $\bar{B}_s^0\to J/\psi \phi$, ``$\bullet$" denotes the effective vertex for the operator insertion.} \label{fig:Fyn}
\end{figure}

We use the form factors obtained in this work to calculate the time-dependent decay width and branching ratio of $\bar{B}^0_s\to J\psi\phi$ mode. For simplicity in checking whether the form factors obtained in this work can give predictions consistent with experiment, we only calculate the branching ratio in naive factorization approach here. The Feynman diagrams for $\bar{B}_s^0\to J/\psi \phi$ decay are shown in Fig.\ref{fig:Fyn}.

 The effective amplitude of $\bar{B}_s^0\to J/\psi \phi$ is
\begin{equation}\label{1}
\mathcal{\bar{A}}_{eff}=\langle J/\psi\phi|\mathcal{H}_{eff}|\bar{B}_s^0\rangle,
\end{equation}
where the effective Hamiltonian is
$$\mathcal{H}_{eff}=\frac{G_F}{\sqrt{2}}\left[V_{cb}V_{cs}^*(C_1O_1+C_2O_2)-V_{tb}V_{ts}^*(\sum_{j=3}^{10}{C_jO_j})\right],$$
with
$$O_1=(\bar{c}_ib_j)_{V-A}(\bar{s}_jc_i)_{V-A},~~O_2=(\bar{c}_ib_i)_{V-A}(\bar{s}_jc_j)_{V-A},$$
$$O_3=(\bar{s}_ib_i)_{V-A}(\bar{c}_jc_j)_{V-A},~~O_4=(\bar{s}_ib_j)_{V-A}(\bar{c}_jc_i)_{V-A},$$
$$O_5=(\bar{s}_ib_i)_{V-A}(\bar{c}_jc_j)_{V+A},~~O_6=(\bar{s}_ib_j)_{V-A}(\bar{c}_jc_i)_{V+A},$$
$$O_7=\frac{3}{2}(\bar{s}_ib_i)_{V-A}\frac{2}{3}(\bar{c}_jc_j)_{V+A},~~O_8=\frac{3}{2}(\bar{s}_ib_j)_{V-A}\frac{2}{3}(\bar{c}_jc_i)_{V+A},$$
$$O_9=\frac{3}{2}(\bar{s}_ib_i)_{V-A}\frac{2}{3}(\bar{c}_jc_j)_{V-A},~~O_{10}=\frac{3}{2}(\bar{s}_ib_j)_{V-A}\frac{2}{3}(\bar{c}_jc_i)_{V-A}.$$

For Wilson coefficient $C_i(\mu)$, we take the value calculated by naive dimensional regularization(NDR) scheme up to the next-to-leading-order at renormalization scale $\mu=m_b$ as \cite{R10}
$$C_1=-0.176;~C_2=1.078;~C_3=0.014;$$
$$C_4=-0.034;~C_5=0.008;~C_6=-0.039;$$
$$C_7=-0.011\alpha;~C_8=0.055\alpha;~C_9=-1.341 \alpha;~C_{10}=0.264\alpha$$
where $\alpha$ is the electromagnetic coupling constant, which takes $\alpha=7.297\times10^{-3}$.

We can divide the total effective amplitude into three parts
\begin{equation}\label{2}
\mathcal{\bar{A}}_{eff}=\mathcal{\bar{A}}_1+\mathcal{\bar{A}}_2+\mathcal{\bar{A}}_3,
\end{equation}
where $\mathcal{\bar{A}}_1$ denotes the contribution of the two tree diagrams of Fig.\ref{fig:Fyn} (a) and (b),
 $\mathcal{\bar{A}}_2$ the contribution of Fig.\ref{fig:Fyn} (c), and $\mathcal{\bar{A}}_3$ the contribution of Fig.\ref{fig:Fyn} (d).

The amplitude of the two tree diagrams is 
\begin{equation}\label{3}
\begin{split}
\mathcal{\bar{A}}_1=
&\frac{G_F}{\sqrt{2}}\left[V_{cb}V_{cs}^*(C_1+\frac{C_2}{N_c})-V_{tb}V_{ts}^*(C_3+\frac{C_4}{N_c}+C_5+\frac{C_6}{N_c}+C_7+\frac{C_8}{N_c}+C_9+\frac{C_{10}}{N_c})\right]\\
&\langle J/\psi|\bar{c}\gamma^\nu (1-\gamma_5)c|0 \rangle\langle \phi|\bar{s}\gamma_\nu (1-\gamma_5)b|\bar{B}_s^0 \rangle,
\end{split}
\end{equation}
while the amplitudes of the two penguin diagrams are
\begin{equation}\label{4}
\begin{split}
\mathcal{\bar{A}}_2=&\frac{G_F}{\sqrt{2}}C_1\left[V_{ub}V_{us}^*\frac{Q_u^2\alpha}{\pi}\int_{0}^{1}{2\bar{x}x(1+\ln{\frac{a^2}{\mu^2}})}dx+
V_{cb}V_{cs}^*\frac{Q_c^2\alpha}{\pi}\int_{0}^{1}{2\bar{x}x(1+\ln{\frac{b^2}{\mu^2}})}dx\right]\\
&\langle J/\psi|\bar{c}\gamma^\nu (1-\gamma_5)c|0 \rangle\langle \phi|\bar{s}\gamma_\nu (1-\gamma_5)b|\bar{B}_s^0 \rangle,
\end{split}
\end{equation}
and
\begin{equation}\label{5}
\begin{split}
\mathcal{\bar{A}}_3=&\frac{G_F}{\sqrt{2}}C_1\left[V_{ub}V_{us}^*\frac{Q_u^2\alpha}{\pi}\int_{0}^{1}{2\bar{x}x\ln{\frac{a^2}{\mu^2}}}dx+
V_{cb}V_{cs}^*\frac{Q_c^2\alpha}{\pi}\int_{0}^{1}{2\bar{x}x\ln{\frac{b^2}{\mu^2}}}dx\right]\\
&\langle J/\psi|\bar{c}\gamma^\nu (1-\gamma_5)c|0 \rangle\langle \phi|\bar{s}\gamma_\nu (1-\gamma_5)b|\bar{B}_s^0 \rangle.
\end{split}
\end{equation}

Because of $\bar{B}_s^0-B_s^0$ mixing, we should consider the decay amplitude of $B_s^0\to J/\psi \phi$ in the analysis of time-dependent decays. Similarly we denote
$\mathcal{A}_{eff}=\langle J/\psi\phi|\mathcal{H}_{eff}|B_s^0\rangle$ and
\begin{equation}\label{6}
\mathcal{A}_{eff}=\mathcal{A}_1+\mathcal{A}_2+\mathcal{A}_3,
\end{equation}
with
\begin{equation}\label{7}
\begin{split}
\mathcal{A}_1=&\frac{G_F}{\sqrt{2}}\left[V_{cb}^*V_{cs}(C_1+\frac{C_2}{N_c})-V_{tb}^*V_{ts}(C_3+\frac{C_4}{N_c}+C_5+\frac{C_6}{N_c}+C_7+\frac{C_8}{N_c}+C_9+\frac{C_{10}}{N_c})\right]\\
&\langle J/\psi|\bar{c}\gamma^\nu (1-\gamma_5)c|0 \rangle\langle \phi|\bar{b}\gamma_\nu (1-\gamma_5)s|B_s^0 \rangle,
\end{split}
\end{equation}
\begin{equation}\label{8}
\begin{split}
\mathcal{A}_2=&\frac{G_F}{\sqrt{2}}C_1\left[V_{ub}^*V_{us}\frac{Q_u^2\alpha}{\pi}\int_{0}^{1}{2\bar{x}x(1+\ln{\frac{a^2}{\mu^2}})}dx+
V_{cb}^*V_{cs}\frac{Q_c^2\alpha}{\pi}\int_{0}^{1}{2\bar{x}x(1+\ln{\frac{b^2}{\mu^2}})}dx\right]\\
&\langle J/\psi|\bar{c}\gamma^\nu (1-\gamma_5)c|0 \rangle\langle \phi|\bar{b}\gamma_\nu (1-\gamma_5)s|B_s^0 \rangle,
\end{split}
\end{equation}
\begin{equation}\label{9}
\begin{split}
\mathcal{A}_3=&\frac{G_F}{\sqrt{2}}C_1\left[V_{ub}^*V_{us}\frac{Q_u^2\alpha}{\pi}\int_{0}^{1}{2\bar{x}x\ln{\frac{a^2}{\mu^2}}}dx+
V_{cb}^*V_{cs}\frac{Q_c^2\alpha}{\pi}\int_{0}^{1}{2\bar{x}x\ln{\frac{b^2}{\mu^2}}}dx\right]\\
&\langle J/\psi|\bar{c}\gamma^\nu (1-\gamma_5)c|0 \rangle\langle \phi|\bar{b}\gamma_\nu (1-\gamma_5)s|B_s^0 \rangle,
\end{split}
\end{equation}
where $\bar{x}=1-x$, $a^2=m_u^2-x(1-x)q^2$, $b^2=m_c^2-x(1-x)q^2$, and $q$ is the transition momentum. $G_F$ is the Fermi constant, $N_c=3$ the color quantum number of quarks, $Q_q(q=u,c)$ the charge of relevant quarks, $C_i(i=1,2,...,10.)$ the Wilson coefficients, and $V_{qb},V_{qs}\;(q=u,c,t)$ the relevant CKM matrix elements, respectively.

There are three polarization states for
$\phi$ meson: one longitudinal state and two transverse polarization
states (right-handed and left-handed).
We define
\begin{equation}\label{41}
h_\lambda~\equiv \langle J/\psi|\bar{c}\gamma^\nu (1-\gamma_5)c|0 \rangle\langle \phi|\bar{s}\gamma_\nu (1-\gamma_5)b|\bar{B}_s^0 \rangle,
\end{equation}
then using Eq. (\ref{10}) and the following matrix elements for $J/\psi$ meson
\begin{eqnarray}
 &&   \langle 0|\bar{c}\gamma_\mu c|{J/\psi}\rangle =m_{J/\psi} f_{J/\psi}
  \varepsilon_\mu^{(\lambda)}, \nonumber \\
&&\langle 0|\bar{c}\gamma_\mu\gamma_5 c|J/\psi\rangle =0,
\end{eqnarray}
we can obtain, for the longitudinal polarization states of the vector mesons
\begin{eqnarray}\label{42}
h_0~=\frac{if_{J/\psi}}{2m_{\phi}}\left[(m_{B_s}^2-m_{\phi}^2-{m_{J/\psi}}^2)(m_{B_s}+m_{\phi})A_1(q^2)-\frac{4{m_{B_s}}^2{p_\phi}^2}{m_{B_s}+m_\phi}A_2(q^2)\right],
\end{eqnarray}
and for the transverse polarization states
\begin{equation}\label{43}
h_\pm~=i{f_{J/\psi}}{m_{J/\psi}}\left[(m_{B_s}+m_{\phi})A_1(q^2)\mp\frac{2m_{B_s}p_\phi V(q^2)}{m_{B_s}+m_{\phi}}\right].
\end{equation}
where ${p_\phi}=\frac{1}{2m_{B_s}}\sqrt{\left[m_{B_s}^2-(m_{\phi}+m_{J/\psi}^2)\right]\left[m_{B_s}^2-(m_{\phi}-m_{J/\psi}^2)\right]}$ is the momentum of $\phi$ meson in the rest frame of $B_s$.

We can write $|\mathcal{\bar{A}}_{eff}|^2$ in terms of the sum of one longitudinal and two transverse polarization amplitudes squared
\begin{equation}\label{44}
|\mathcal{\bar{A}}_{eff}|^2=|(\mathcal{\bar{A}}_{eff})_L|^2+|(\mathcal{\bar{A}}_{eff})_+|^2+|(\mathcal{\bar{A}}_{eff})_-|^2,
\end{equation}
where
\begin{equation}\label{45}
\begin{split}
|(\mathcal{\bar{A}}_{eff})_L|^2=|(\mathcal{\bar{A}}_1)_L|^2+|(\mathcal{\bar{A}}_2)_L|^2+|(\mathcal{\bar{A}}_3)_L|^2,\\
|(\mathcal{\bar{A}}_{eff})_\pm|^2=|(\mathcal{\bar{A}}_1)_\pm|^2+|(\mathcal{\bar{A}}_2)_\pm|^2+|(\mathcal{\bar{A}}_3)_\pm|^2.
\end{split}
\end{equation}

In the same way, $|\mathcal{A}_{eff}|^2$ can be written as
\begin{equation}\label{46}
|\mathcal{A}_{eff}|^2=|(\mathcal{A}_{eff})_L|^2+|(\mathcal{A}_{eff})_+|^2+|(\mathcal{A}_{eff})_-|^2,
\end{equation}
where
\begin{equation}\label{47}
\begin{split}
|(\mathcal{A}_{eff})_L|^2=|(\mathcal{A}_1)_L|^2+|(\mathcal{A}_2)_L|^2+|(\mathcal{A}_3)_L|^2,\\
|(\mathcal{A}_{eff})_\pm|^2=|(\mathcal{A}_1)_\pm|^2+|(\mathcal{A}_2)_\pm|^2+|(\mathcal{A}_3)_\pm|^2.
\end{split}
\end{equation}
The relevant expressions for the terms in the right hand of Eqs.(\ref{45}) and (\ref{47}) are given in the Appendix C. Then we obtain the transverse time-dependent decay width
\begin{eqnarray}
\begin{split}\label{48}
\Gamma(\bar{B}_s^0(t)\to J/\psi \phi)_T^\pm~=&\frac{p_\phi}{8\pi m_{B_s}^2}\frac{1}{2}e^{-\Gamma_{B_s}t}\left[(|(\mathcal{A}_{eff})_\pm|^2+|(\mathcal{\bar{A}}_{eff})_\pm|^2)
\cosh{\frac{\Delta\Gamma}{2}t}\right. \\
&\left.-(|(\mathcal{A}_{eff})_\pm|^2-|(\mathcal{\bar{A}}_{eff})_\pm|^2)\cos{{\Delta m}t}\right. \\
&\left.+2Re(\frac{p}{q}(\mathcal{A}_{eff})_\pm({\mathcal{\bar{A}}_{eff}}^*)_\pm)\sinh{\frac{\Delta\Gamma}{2}t}
-2Im(\frac{p}{q}(\mathcal{A}_{eff})_\pm({\mathcal{\bar{A}}_{eff}}^*)_\pm)\sin{{\Delta m}t}\right],
\end{split}
\end{eqnarray}
and the longitudinal time-dependent decay width
\begin{eqnarray}\label{49}
\begin{split}
\Gamma(\bar{B}_s^0(t)\to J/\psi \phi)_L~=&\frac{p_\phi}{8\pi m_{B_s}^2}\frac{1}{2}e^{-\Gamma_{B_s}t}\left[(|(\mathcal{A}_{eff})_L|^2+|(\mathcal{\bar{A}}_{eff})_L|^2)
\cosh{\frac{\Delta\Gamma}{2}t}\right.\\
&\left.-(|(\mathcal{A}_{eff})_L|^2-|(\mathcal{\bar{A}}_{eff})_L|^2)\cos{{\Delta m}t}\right. \\
&\left.+2Re(\frac{p}{q}(\mathcal{A}_{eff})_L({\mathcal{\bar{A}}_{eff}}^*)_L)\sinh{\frac{\Delta\Gamma}{2}t}
-2Im(\frac{p}{q}(\mathcal{A}_{eff})_L({\mathcal{\bar{A}}_{eff}}^*)_L)\sin{{\Delta m}t}\right].
\end{split}
\end{eqnarray}

We take $\frac{p}{q}=\frac{V_{tb}V_{ts}^*}{V_{tb}^*V_{ts}}=e^{-i2\beta_s}$, $\beta_s=0.0185$, $\frac{\Delta\Gamma}{\Gamma_{B_s}}=0.122$, $\frac{\Delta m}{\Gamma_{B_s}}=26.79$ and the total decay width of $B_s$ meson is $\Gamma_{B_s}=4.362\times10^{-13}\;\mbox{GeV}$ \cite{R17}.
Finally, the combined transverse and total time-dependent decay widths are
\begin{equation}\label{50}
\begin{split}
\Gamma(\bar{B}_s^0(t)\to J/\psi \phi)_T=\Gamma(\bar{B}_s^0(t)\to J/\psi \phi)_T^++\Gamma(\bar{B}_s^0(t)\to J/\psi \phi)_T^-,\\
\Gamma(\bar{B}_s^0(t)\to J/\psi \phi)=\Gamma(\bar{B}_s^0(t)\to J/\psi \phi)_L+\Gamma(\bar{B}_s^0(t)\to J/\psi \phi)_T.
\end{split}
\end{equation}

And the total time-dependent decay width is \cite{R17}
\begin{eqnarray}
\begin{split}\label{51}
\Gamma(\bar{B}_s^0(t)\to J/\psi \phi)~=&\frac{p_\phi}{8\pi m_{B_s}^2}\frac{1}{2}e^{-\Gamma_{B_s}t}\left[(|\mathcal{A}_{eff}|^2+|\mathcal{\bar{A}}_{eff}|^2)
\cosh{\frac{\Delta\Gamma}{2}t}\right.\\
&\left.-(|\mathcal{A}_{eff}|^2-|\mathcal{\bar{A}}_{eff}|^2)\cos{{\Delta m}t}+2Re(\frac{p}{q}\mathcal{A}_{eff}{\mathcal{\bar{A}}_{eff}}^*)\sinh{\frac{\Delta\Gamma}{2}t}\right.\\
&\left.-2Im(\frac{p}{q}\mathcal{A}_{eff}{\mathcal{\bar{A}}_{eff}}^*)\sin{{\Delta m}t}\right].
\end{split}
\end{eqnarray}
Similarly, we can also obtain the total time-dependent decay width of $B_s^0(t)\to J/\psi \phi$, which is \cite{R17}
\begin{eqnarray}
\begin{split}\label{52}
\Gamma(B_s^0(t)\to J/\psi \phi)~=&\frac{p_\phi}{8\pi m_{B_s}^2}\frac{1}{2}e^{-\Gamma_{B_s}t}\left[(|\mathcal{A}_{eff}|^2+|\mathcal{\bar{A}}_{eff}|^2)
\cosh{\frac{\Delta\Gamma}{2}t}\right.\\
&\left.+(|\mathcal{A}_{eff}|^2-|\mathcal{\bar{A}}_{eff}|^2)\cos{{\Delta m}t}+2Re(\frac{q}{p}{\mathcal{A}_{eff}}^*\mathcal{\bar{A}}_{eff})\sinh{\frac{\Delta\Gamma}{2}t}\right.\\
&\left.-2Im(\frac{q}{p}{\mathcal{A}_{eff}}^*\mathcal{\bar{A}}_{eff})\sin{{\Delta m}t}\right].
\end{split}
\end{eqnarray}
Integrating the above time-dependent decay widths over $t$ from zero to infinity, we can get the relevant branching ratios \cite{Duni2001,Bruyn2012}
\begin{equation}\label{53}
\begin{split}
Br(B_s\to J/\psi \phi)_T=\frac{1}{2}\int_{0}^{\infty}\left[{\Gamma(\bar{B}_s^0(t)\to J/\psi \phi)_T}+\Gamma(B_s^0(t)\to J/\psi \phi)_T\right]dt,\\
Br(B_s\to J/\psi \phi)_L=\frac{1}{2}\int_{0}^{\infty}\left[{\Gamma(\bar{B}_s^0(t)\to J/\psi \phi)_L}+\Gamma(B_s^0(t)\to J/\psi \phi)_L\right]dt,\\
Br(B_s\to J/\psi \phi)_{total}=\frac{1}{2}\int_{0}^{\infty}\left[{\Gamma(\bar{B}_s^0(t)\to J/\psi \phi)}+\Gamma(B_s^0(t)\to J/\psi \phi)\right]dt.
\end{split}
\end{equation}

Substituting the values for the relevant parameters and quantities into the above equation we can get
\begin{equation}\label{54}
\begin{split}
Br(B_s\to J/\psi \phi)_L=(0.42\pm 0.17)\times10^{-3},\\
Br(B_s\to J/\psi \phi)_T=(0.50\pm 0.09)\times10^{-3},
\end{split}
\end{equation}
and the total total decay branching ratio is
\begin{equation}\label{55}
Br(B_s\to J/\psi \phi)=(0.92\pm 0.26)\times10^{-3},
\end{equation}
which are in good agreement with experimental data within uncertainties \cite{R17}:
$$Br(B_s\to J/\psi \phi)_L^{exp}=(0.56\pm 0.04)\times10^{-3},$$
$$Br(B_s\to J/\psi \phi)_T^{exp}=(0.52\pm 0.04)\times10^{-3},$$
$$Br(B_s\to J/\psi \phi)^{exp}=(1.08\pm 0.08)\times10^{-3}.$$

\section{Summary}

We calculate the $\bar{B}^0_s\to\phi$ transition form factors by QCD sum rule method.
The form factors are expressed in terms of two Borel parameters $M_1^2$,
$M_2^2$ and relevant Borel transformation coefficients.
We take the two Borel parameters $M_1^2$, $M_2^2$ as independent parameters
and find the ``stable windows" in the two-dimensional area of $M_1^2$ and $M_2^2$ for
the transition form factors $V$, $A_0$, $A_1$ and $A_2$. Our results are compatible
with that obtained by LCSR and CQM methods in the literature. Finally, we apply
the results of the transition form factors $V$, $A_0$, $A_1$ and $A_2$ to the nonleptonic decay
process of $\bar{B}_s^0\to J/\psi \phi$. We calculate the branching ratios for all the possible
polarization states of the vector mesons. The branching ratios we obtained are well consistent
with experimental data.

\acknowledgments

This work is supported in part by the National Natural Science
Foundation of China under Contracts No. 11875168 and No. 11375088.

\newpage

\begin{center}{\bf Appendix A }\end{center}

Some basic formulas are needed to perform the integration in Eq.~(\ref{25}) are given here.

$$
\begin{array}{ll}
I=\int d^4k \delta (k^2-m_0^2)\delta[(k+p_1)^2-m_1^2]\delta
[(k+p_2)^2-m_2^2]=\frac{\pi}{2\sqrt{\lambda}},
\end{array}
\eqno(A1)$$

$$
\begin{array}{ll}
I_\mu =\int d^4k k_\mu \delta
(k^2-m_0^2)\delta[(k+p_1)^2-m_1^2]\delta \
[(k+p_2)^2-m_2^2]\nonumber \\
\equiv a_1 p_{1\mu}+b_1 p_{2\mu},\label{27}\\[4mm]
\left\{ \begin{array}{ll}
          a_1~=&-\frac{\pi}{2\lambda ^{3/2}}[s_2(-s_1+s_2-q^2)+
            (s_1+s_2-q^2)(m_0^2-m_2^2)\\[3mm]
             &~-2s_2(m_0^2-m_1^2)],\\[3mm]
          b_1~=&-\frac{\pi}{2\lambda ^{3/2}}[s_1(-s_2+s_1-q^2)+
            (s_1+s_2-q^2)(m_0^2-m_1^2)\\[3mm]
              &~-2s_1(m_0^2-m_2^2)],
          \end{array}\right. \nonumber
\end{array}
\eqno(A2)$$

$$
\begin{array}{ll}
I_{\mu\nu}=\int d^4k k_\mu k_\nu\delta
(k^2-m_0^2)\delta[(k+p_1)^2-m_1^2]\delta \
[(k+p_2)^2-m_2^2]\nonumber
\\ \equiv a_2p_{1\mu}p_{1\nu}+b_2p_{2\mu\nu}+c_2(p_{1\mu}p_{2\nu}+
          p_{1\nu}p_{2\mu})+d_2g_{\mu\nu},\label{28}\\[4mm]
\left\{ \begin{array}{ll}
      D_1~\equiv & s_1-m_1^2+m_0^2,~~~~~D_2\equiv s_2-m_2^2+m_0^2,\\[3mm]
      a_2~=& \frac{\pi}{\lambda^{3/2}}m_0^2s_2
        +\frac{1}{\lambda}[3s_2D_1a_1-(s_1+s_2-q^2)D_2b_1+s_2D_2b_1],\\[3mm]
      b_2~=& \frac{\pi}{\lambda^{3/2}}m_0^2s_1
        +\frac{1}{\lambda}[s_1D_1a_1-(s_1+s_2-q^2)D_1b_1+3s_1D_2b_1],\\[3mm]
      c_2~=&-\frac{\pi}{\lambda^{3/2}}m_0^2\frac{1}{2}(s_1+s_2-q^2)\\[3mm]
         &-\frac{1}{\lambda}[\frac{1}{2}(s_1+s_2-q^2)D_1a_1-2s_2D_1b_1
         +\frac{3}{2}(s_1+s_2-q^2)D_2b_1],\\[3mm]
      d_2~=&\frac{\pi}{4\sqrt{\lambda}}+\frac{1}{4}[D_1a_1+D_2b_1].
          \end{array} \right. \nonumber
\end{array}
\eqno(A3)$$
where $\lambda (s_1,s_2,q^2)=(s_1+s_2-q^2)^2-4s_1s_2$.

\begin{center}{\bf Appendix B }\end{center}

The results of relevant Borel transformed Coefficients for the transition form factors in Eq.~(\ref{21}) are given here.

\noindent 1) Borel transformed $f_0$:

$$ \hat{B}f_0= \hat{B}f_0^{pert}+ \hat{B}f_0^{(3)}+
\hat{B}f_0^{(5)}+ \hat{B}f_0^{(6)}\; ,$$
where,
$$
\begin{array}{ll}
 \hat{B}f_0^{pert}~=&\int^{s_2^0}_{4m_s^2}
 ds_2\int^{s_1^0}_{s_1^L}ds_1
\displaystyle
\frac{3e^{-s_1/M_1^2-s_2/M_2^2}}{4M_1^2M_2^2\pi^2\lambda^{3/2}}[-{s_2} m_b(2 m_s^2+q^2\\
&+{s_1}-{s_2})-2{s_2} m_b^2 m_s+2{s_2} m_b^3+m_s(\lambda +2{s_2} m_s^2\\
&+q^2{s_2}+{s_1}{s_2}-{s_2}^2)],
\end{array}
\eqno(B1)$$
where $\lambda=(s_1+s_2-q^2)^2-4s_1 s_2$. The lower limit of the integration
$s_1^L$  is determined by requiring that all internal quarks
are on their mass shell \cite{R1}
$$ s_1^L=\frac{m_b^2}{m_b^2-q^2}s_2+m_b^2\; ,$$
and
$$
\begin{array}{ll}
 \hat{B}f_0^{(3)}~=&-\displaystyle
\frac{e^{-m_b^2/M_1^2-m_s^2/M_2^2}}{6M_1^8M_2^8}[{M_1}^2{M_2}^2 m_b^2 m_s^2 ({M_1}^2+{M_2}^2) (3{M_2}^2-m_s^2)\\
&+{M_1}^2{M_2}^2 m_b m_s({M_2}^2 m_s^2({M_1}^2+{M_2}^2+q^2)-m_s^4 ({M_1}^2+{M_2}^2)\\
&-3{M_1}^2{M_2}^4)-{M_2}^4 m_b^3 m_s^3({M_1}^2+{M_2}^2)+{M_1}^4(-3{M_2}^4 m_s^2 ({M_1}^2\\
&+q^2)+{M_2}^2 m_s^4(4{M_1}^2+4{M_2}^2+q^2)-m_s^6({M_1}^2+{M_2}^2)\\
&+6{M_1}^2{M_2}^6)]
\times\langle \bar{s}s\rangle\; ,
 \end{array}
\eqno(B2)
$$

$$
\begin{array}{ll}
 \hat{B}f_0^{(5)}~=&-\displaystyle
\frac{e^{-m_b^2/M_1^2-m_s^2/M_2^2}}{12M_1^8M_2^8}[{M_1}^2{M_2}^2 m_b m_s(m_s^2({M_1}^2+{M_2}^2)\\
&+{M_2}^2(-2{M_1}^2+2{M_2}^2-q^2))-{M_1}^2{M_2}^2 m_b^2({M_1}^2\\
&+{M_2}^2)(3{M_2}^2-m_s^2)+{M_2}^4 m_b^3 m_s({M_1}^2
+{M_2}^2)\\
&+{M_1}^4(-{M_2}^2 m_s^2(5{M_1}^2+3{M_2}^2+q^2)+m_s^4({M_1}^2+{M_2}^2)\\
&+{M_2}^4(3({M_2}^2+q^2)-{M_1}^2))]\times g\langle
\bar{s}\sigma TG s \rangle\; ,
\end{array}
\eqno(B3)
$$

$$
\begin{array}{ll}
 \hat{B}f_0^{(6)}~=&\displaystyle
\frac{e^{-m_b^2/M_1^2-m_s^2/M_2^2}}{81M_1^8M_2^8(m_b^2-q^2)m_s^3}[{M_1}^2{M_2}^2 m_b^4 m_s^4({M_1}^2+{M_2}^2)\\
&+{M_2}^4 m_b^5 m_s^3({M_1}^2+{M_2}^2)+{M_2}^2 m_b^3 m_s^3({M_1}^2 m_s^2({M_1}^2+{M_2}^2)\\
&-{M_2}^2(-2{M_1}^4+{M_1}^2(13{M_2}^2+2q^2)+{M_2}^2q^2))\\
&+{M_1}^2{M_2}^2 m_b m_s(36{M_1}^4{M_2}^4 (e^{\frac{m_s^2}{{M_2}^2}}-1)+{M_2}^2q^2 m_s^2 (-2{M_1}^2\\
&+13{M_2}^2+q^2)-q^2m_s^4({M_1}^2+{M_2}^2))+m_b^2(54{M_1}^6{M_2}^4 m_s^2\\
&-54{M_1}^6 {M_2}^6(e^{\frac{m_s^2}{{M_2}^2}}-1)
-{M_1}^2{M_2}^2 m_s^4({M_1}^4+2{M_1}^2(5{M_2}^2+q^2)\\
&+{M_2}^2q^2){M_1}^4 m_s^6({M_1}^2+{M_2}^2))
+{M_1}^4({M_2}^2q^2 m_s^4({M_1}^2+10{M_2}^2\\
&+q^2)-q^2 m_s^6({M_1}^2+{M_2}^2)+54{M_1}^2{M_2}^6q^2(e^{\frac{m_s^2}{{M_2}^2}}-1)\\
&-18{M_1}^2{M_2}^4 m_s^2({M_2}^2 (e^{\frac{m_s^2}{{M_2}^2}}-1)+3q^2)] \times g^2\langle \bar{s}s\rangle^2\; .
\end{array}
\eqno(B4)
$$

\noindent 2) Borel transformed result for $f_1+f_3$ :

$$ \hat{B}(f_1+f_3)= \hat{B}f_+^{pert}+ \hat{B}f_+^{(3)}+
\hat{B}f_+^{(5)}+ \hat{B}f_+^{(6)}\; ,$$
where
$$
\begin{array}{ll}
 \hat{B}f_+^{pert}~=&\int^{s_2^0}_{4m_s^2}
 ds_2\int^{s_1^0}_{s_1^L}ds_1
\displaystyle
\frac{3e^{-s_1/M_1^2-s_2/M_2^2}}{4M_1^2M_2^2\pi^2\lambda^{5/2}}
\{2{s_2} m_b^3(2q^2(3 m_s^2+{s_1}+{s_2})+2q^4\\
&+\lambda -6 m_s^2 ({s_1}-{s_2})-4{s_1}^2+8{s_1}{s_2}-4{s_2}^2)+2 {s_2} m_b^2 m_s(-2q^2(3 m_s^2+{s_1}+{s_2})\\
&-2q^4-3 \lambda +6 m_s^2 ({s_1}-{s_2})+4{s_1}^2-8{s_1}{s_2}+4 {s_2}^2)+m_b(q^2(-2 m_s^2(\lambda +2{s_1}{s_2}+2{s_2}^2)\\
&-6{s_2} m_s^4+{s_2}(-\lambda +2{s_1}^2-6{s_1}{s_2}+4 {s_2}^2))-2{s_2}q^4(2 m_s^2+2{s_1}+{s_2})+2 m_s^2(4{s_1}^2{s_2}\\
&+\lambda{s_1}-8{s_1}{s_2}^2+4{s_2}^3-2\lambda  {s_2})+6{s_2} m_s^4 ({s_1}-{s_2})+{s_2} ({s_1}-{s_2})(-\lambda +2{s_1}^2\\
&-4{s_1}{s_2}+2{s_2}^2))+6{s_2}m_b^4 m_s (q^2-{s_1}+{s_2})-6{s_2} m_b^5(q^2-{s_1}+{s_2})\\
&+m_s(q^2(2 m_s^2(\lambda +2{s_1}{s_2}+2{s_2}^2)+6{s_2} m_s^4+{s_2} (3\lambda -2{s_1}^2+6{s_1}{s_2}-4{s_2}^2))\\
&+2{s_2}q^4(2 m_s^2+2{s_1}+{s_2})+\lambda ^2-2 m_s^2(4{s_1}^2 {s_2}+\lambda{s_1}-8{s_1}{s_2}^2+4{s_2}^3-4 \lambda{s_2})\\
&+6{s_2} m_s^4({s_2}-{s_1})-2{s_1}^3{s_2}+6{s_1}^2 {s_2}^2-6{s_1}{s_2}^3+3\lambda{s_1}{s_2}+2{s_2}^4-3\lambda{s_2}^2)\}\; ,
\end{array}
\eqno(B5)$$

$$
\begin{array}{ll}
 \hat{B}f_+^{(3)}~=&-\displaystyle
\frac{e^{-m_b^2/M_1^2-m_s^2/M_2^2}}{6M_1^8M_2^8}\{-{M_1}^2{M_2}^2 m_b m_s({M_2}^2 m_s^2(-q^2+{M_1}^2+{M_2}^2)\\
&+m_s^4({M_1}^2+{M_2}^2)+3{M_1}^2{M_2}^4)+{M_1}^4({M_2}^2 m_s^4(q^2+4 ({M_1}^2+{M_2}^2))\\
&-3{M_2}^4 m_s^2(q^2+{M_1}^2-2{M_2}^2)-m_s^6({M_1}^2+{M_2}^2)+6{M_1}^2{M_2}^6)\\
&+{M_1}^2{M_2}^2 m_b^2 m_s^2({M_1}^2+{M_2}^2)(3{M_2}^2-m_s^2)-{M_2}^4 m_b^3 m_s^3({M_1}^2\\
&+{M_2}^2)\}
\times\langle \bar{s}s\rangle\; ,
 \end{array}
\eqno(B6)$$

$$\begin{array}{ll}
 \hat{B}f_+^{(5)}~=&\displaystyle
\frac{e^{-m_b^2/M_1^2-m_s^2/M_2^2}}{12M_1^8M_2^8}\{
q^2({M_1}^2{M_2}^4 m_b m_s+{M_1}^4{M_2}^2(m_s^2-3{M_2}^2))\\
&-{M_1}^2{M_2}^2 m_b m_s(m_s^2({M_1}^2+{M_2}^2)+2{M_2}^2 ({M_1}^2+2{M_2}^2))\\
&+{M_1}^2{M_2}^2 m_b^2({M_1}^2+{M_2}^2)(3{M_2}^2-m_s^2)-{M_2}^4 m_b^3 m_s ({M_1}^2+{M_2}^2)\\
&+{M_1}^4(-m_s^4({M_1}^2+{M_2}^2)+m_s^2(5{M_1}^2{M_2}^2+7{M_2}^4)\\
&+{M_2}^4({M_1}^2+3 {M_2}^2))\}\times g\langle
\bar{s}\sigma TG s \rangle\; ,
\end{array}
\eqno(B7)$$

\newpage
$$
\begin{array}{ll}
 \hat{B}f_+^{(6)}~=&\displaystyle
\frac{e^{-m_b^2/M_1^2-m_s^2/M_2^2}}{81M_1^8M_2^8(m_b^2-q^2)m_s^3}
\{{M_1}^2{M_2}^2q^4 m_s^3({M_2}^2 m_b+{M_1}^2 m_s)\\
&+q^2(-{M_1}^2{M_2}^2 m_b m_s^5 ({M_1}^2+{M_2}^2)+{M_1}^2{M_2}^2 m_s^4(-m_b^2(2{M_1}^2+{M_2}^2)\\
&+{M_1}^4+10{M_1}^2{M_2}^2)-{M_2}^4 m_b m_s^3(m_b^2(2{M_1}^2+{M_2}^2)+4{M_1}^4-11 {M_1}^2{M_2}^2)\\
&-54{M_1}^6{M_2}^4 m_s^2+54{M_1}^6{M_2}^6(e^{\frac{m_s^2}{{M_2}^2}}-1)-{M_1}^4 m_s^6 ({M_1}^2+{M_2}^2))\\
&+{M_1}^2 {M_2}^2 m_b^4 m_s^4({M_1}^2+{M_2}^2)+{M_2}^4 m_b^5 m_s^3 ({M_1}^2+{M_2}^2)\\
&+{M_1}^2 {M_2}^2 m_b^3 m_s^3(m_s^2({M_1}^2+{M_2}^2)+4{M_1}^2{M_2}^2-11{M_2}^4)\\
&+{M_1}^4 m_b^2(54{M_1}^2{M_2}^4 m_s^2-{M_2}^2 m_s^4({M_1}^2+10{M_2}^2)+m_s^6({M_1}^2+{M_2}^2)\\
&-54{M_1}^2{M_2}^6 (e^{\frac{m_s^2}{{M_2}^2}}-1))-18 {M_1}^6{M_2}^6 m_s^2(e^{\frac{m_s^2}{{M_2}^2}}-1)\}
 \times g^2\langle \bar{s}s\rangle^2\; .
\end{array}
\eqno(B8)$$

\noindent 3) Borel transformed result for $f_1-f_3$ :
$$ \hat{B}(f_1-f_3)= \hat{B}f_-^{pert}+ \hat{B}f_-^{(3)}+
\hat{B}f_-^{(5)}+ \hat{B}f_-^{(6)}\; ,$$
where
$$
\begin{array}{ll}
 \hat{B}f_-^{pert}~=&\int^{s_2^0}_{4m_s^2}
 ds_2\int^{s_1^0}_{s_1^L}ds_1
\displaystyle
\frac{-3e^{-s_1/M_1^2-s_2/M_2^2}}{4M_1^2M_2^2\pi^2\lambda^{5/2}}
\{2{s_2} m_b^3(2q^2(3 m_s^2+{s_1}-5{s_2})+2q^4\\
&+\lambda -6 m_s^2 ({s_1}+3{s_2})-4{s_1}^2-4{s_1}{s_2}+8 {s_2}^2)+2 {s_2} m_b^2 m_s(-2q^2(3 m_s^2\\
&+{s_1}-5{s_2})-2q^4+\lambda +6 m_s^2 ({s_1}+3{s_2})+4{s_1}^2+4{s_1}{s_2}-8 {s_2}^2)\\
&+m_b(q^2(-2 m_s^2(\lambda +2{s_1}{s_2}-10{s_2}^2)-6{s_2} m_s^4-{s_2}(\lambda -2{s_1}^2-10{s_1}{s_2}\\
&+4 {s_2}^2))+2{s_2}q^4(-2 m_s^2-2{s_1}+{s_2})+2 m_s^2 ({s_1}+2{s_2})(\lambda +4{s_1}{s_2}-4{s_2}^2)\\
&+6{s_2} m_s^4 ({s_1}+3{s_2})+{s_2}({s_1}-{s_2})(-\lambda +2{s_1}^2-2{s_2}^2))-6{s_2} m_b^4 m_s(-q^2\\
&+{s_1}+3{s_2})+6{s_2} m_b^5 (-q^2+{s_1}+3{s_2})-m_s(q^2(-2 m_s^2(\lambda +2{s_1}{s_2}-10{s_2}^2)\\
&-6{s_2} m_s^4+{s_2}(\lambda +2{s_1}^2+10{s_1} {s_2}-4{s_2}^2))+2{s_2}q^4(-2 m_s^2-2{s_1}+{s_2})-\lambda^2\\
&+2 m_s^2(4{s_1}^2{s_2}+\lambda{s_1}+4{s_1}{s_2}^2-8 {s_2}^3+4 \lambda{s_2})+6{s_2} m_s^4 ({s_1}+3{s_2})+2{s_1}^3{s_2}\\
&-2{s_1}^2{s_2}^2-2{s_1}{s_2}^3+\lambda{s_1} {s_2}+2{s_2}^4-\lambda{s_2}^2)\}\; ,
\end{array}
\eqno(B9)
$$

$$
\begin{array}{ll}
 \hat{B}f_-^{(3)}~=&\displaystyle
\frac{e^{-m_b^2/M_1^2-m_s^2/M_2^2}}{6M_1^8M_2^8}\{-{M_1}^2{M_2}^2 m_b m_s({M_2}^2 m_s^2(-q^2+{M_1}^2-3 {M_2}^2)\\
&+m_s^4({M_1}^2+{M_2}^2)+3{M_1}^2{M_2}^4)+{M_1}^4({M_2}^2 m_s^4(q^2+4 ({M_1}^2+{M_2}^2))\\
&-3{M_2}^4 m_s^2 (q^2+{M_1}^2+2{M_2}^2)-m_s^6({M_1}^2+{M_2}^2)+6{M_1}^2{M_2}^6)\\
&+{M_1}^2{M_2}^2 m_b^2 m_s^2({M_1}^2+{M_2}^2)(3 {M_2}^2-m_s^2)-{M_2}^4 m_b^3 m_s^3({M_1}^2\\
&+{M_2}^2)\}
\times\langle \bar{s}s\rangle\; ,
 \end{array}
\eqno(B10)
$$

$$
\begin{array}{ll}
 \hat{B}f_-^{(5)}~=&-\displaystyle
\frac{e^{-m_b^2/M_1^2-m_s^2/M_2^2}}{12M_1^8M_2^8}\{
q^2({M_1}^2{M_2}^4 m_b m_s+{M_1}^4{M_2}^2(m_s^2-3{M_2}^2))\\
&+{M_1}^2{M_2}^2 m_b^2({M_1}^2+{M_2}^2)(3 {M_2}^2-m_s^2)-{M_2}^4 m_b^3 m_s({M_1}^2+{M_2}^2)\\
&-m_b(2{M_1}^4{M_2}^4 m_s+{M_1}^2{M_2}^2 m_s^3({M_1}^2+{M_2}^2))+{M_1}^4(-m_s^4({M_1}^2\\
&+{M_2}^2)+m_s^2(5{M_1}^2{M_2}^2-{M_2}^4)+{M_2}^4({M_1}^2
-9{M_2}^2))\}\\
&\times g\langle
\bar{s}\sigma TG s \rangle\; ,
\end{array}
\eqno(B11)
$$

$$
\begin{array}{ll}
 \hat{B}f_-^{(6)}~=&-\displaystyle
\frac{e^{-m_b^2/M_1^2-m_s^2/M_2^2}}{81M_1^8M_2^8(m_b^2-q^2)m_s^3}
\{{M_1}^2{M_2}^2q^4 m_s^3({M_2}^2 m_b+{M_1}^2 m_s)\\
&+q^2(-{M_1}^2{M_2}^2 m_b m_s^5 ({M_1}^2+{M_2}^2)+{M_1}^2 {M_2}^2 m_s^4(-m_b^2(2{M_1}^2+{M_2}^2)\\
&+{M_1}^4+10{M_1}^2{M_2}^2)-{M_2}^4 m_b m_s^3(m_b^2(2 {M_1}^2+{M_2}^2)+4 {M1}^4\\
&-15{M_1}^2{M_2}^2)-54{M_1}^6{M_2}^4 m_s^2+54{M_1}^6{M_2}^6 (e^{\frac{m_s^2}{{M_2}^2}}-1)-{M_1}^4 m_s^6 ({M_1}^2\\
&+{M_2}^2))+{M_1}^2{M_2}^2 m_b^4 m_s^4({M_1}^2+{M_2}^2)+{M_2}^4 m_b^5 m_s^3 ({M_1}^2+{M_2}^2)\\
&+{M_1}^2 {M_2}^2 m_b^3 m_s^3(m_s^2({M_1}^2+{M_2}^2)+4{M_1}^2{M_2}^2-15{M_2}^4)\\
&+{M_1}^4 m_b^2(54{M_1}^2{M_2}^4 m_s^2-{M_2}^2 m_s^4({M_1}^2+10{M_2}^2)+m_s^6({M_1}^2\\
&+{M_2}^2)-54{M_1}^2{M_2}^6 (e^{\frac{m_s^2}{{M_2}^2}}-1))+54 {M_1}^6{M_2}^6 m_s^2(e^{\frac{m_s^2}{{M_2}^2}}-1)\}\\
&\times g^2\langle \bar{s}s\rangle^2\; .
\end{array}
\eqno(B12)
$$

\noindent 4) Borel transformed result for $f_5$ :
$$ \hat{B}(f_5)= \hat{B}f_5^{pert}+ \hat{B}f_5^{(3)}+
\hat{B}f_5^{(5)}+ \hat{B}f_5^{(6)}\; ,$$
where,
$$
\begin{array}{ll}
 \hat{B}f_5^{pert}~=&\int^{s_2^0}_{4m_s^2}
 ds_2\int^{s_1^0}_{s_1^L}ds_1
\displaystyle
\frac{-3e^{-s_1/M_1^2-s_2/M_2^2}}{8M_1^2M_2^2\pi^2\lambda^{3/2}}
\{m_b(2{s_2}q^2(m_s^2+{s_1})+2 m_s^2(\lambda\\
&+{s_1}{s_2}-{s_2}^2)+2{s_2} m_s^4+\lambda{s_2})-2{s_2} m_b^3(q^2+2 m_s^2+{s_1}-{s_2})\\
&+2 {s_2} m_b^2 m_s(q^2+2 m_s^2+{s_1}-{s_2})-m_s(q^2(\lambda +2{s_2} m_s^2+2{s_1}{s_2})\\
&+2 m_s^2(\lambda +{s_1}{s_2}-{s_2}^2)+2{s_2} m_s^4-\lambda{s_1})-2{s_2} m_b^4 m_s+2{s_2} m_b^5\} \; ,
\end{array}
\eqno(B13)
$$

\newpage
$$
\begin{array}{ll}
 \hat{B}f_5^{(3)}~=&-\displaystyle
\frac{e^{-m_b^2/M_1^2-m_s^2/M_2^2}}{12M_1^8M_2^8}\{{M_1}^2{M_2}^2 q^4 m_s^2({M_2}^2 m_b m_s+{M_1}^2(m_s^2-3 {M_2}^2))\\
&+q^2 ({M_1}^2 {M_2}^2 m_b^2 m_s^2(3{M_2}^2(2{M_1}^2+{M_2}^2)-m_s^2(2{M_1}^2+3{M_2}^2))\\
&+{M_1}^2{M_2}^2 m_b m_s(-m_s^4(3{M_1}^2+2 {M_2}^2)+m_s^2 (9 {M_1}^2{M_2}^2+2{M_2}^4)\\
&-3{M_1}^2{M_2}^4)-{M_2}^4 m_b^3 m_s^3(2{M_1}^2+{M_2}^2)+{M_1}^4(-m_s^6({M_1}^2+2{M_2}^2)\\
&-3 m_s^2(2{M_1}^2 {M_2}^4+{M_2}^6)+m_s^4(5{M_1}^2{M_2}^2+7{M_2}^4)+6{M_1}^2{M_2}^6))\\
&-{M_2}^2 m_b^4 m_s^2({M_1}^2+{M_2}^2)(3{M_1}^2{M_2}^2-m_s^2 ({M_1}^2+2{M_2}^2))\\
&+{M_2}^4 m_b^5 m_s^3({M_1}^2+{M_2}^2)+m_b^3(3{M_1}^4{M_2}^6 m_s-9{M_1}^2{M_2}^4 m_s^3 ({M_1}^2\\
&+{M_2}^2)+m_s^5 (3 {M_1}^4{M_2}^2+4{M_1}^2{M_2}^4+{M_2}^6))+{M_1}^2 m_b^2(m_s^2(6{M_1}^4{M_2}^4\\
&+9{M_1}^2{M_2}^6)+m_s^6({M_1}^4+4{M_1}^2{M_2}^2+3 {M_2}^4)-m_s^4(5{M_1}^4{M_2}^2\\
&+11{M_1}^2{M_2}^4+6{M_2}^6)-6{M_1}^4{M_2}^6)+{M_1}^2 m_b m_s(m_s^2(11{M_1}^4{M_2}^4\\
&+8{M_1}^2 {M_2}^6)+m_s^6(2{M_1}^4+3{M_1}^2{M_2}^2+{M_2}^4)-m_s^4(11{M_1}^4{M_2}^2\\
&+13{M_1}^2{M_2}^4+2{M_2}^6)-15{M_1}^4 {M_2}^6)+{M_1}^4 m_s^2 (-4 {M_2}^2 m_s^4({M_1}^2\\
&+{M_2}^2)+m_s^6({M_1}^2+{M_2}^2)+m_s^2(5{M_1}^2{M_2}^4+2{M_2}^6)+3{M_1}^2{M_2}^6)\}\\
&\times\langle \bar{s}s\rangle\; ,
 \end{array}
\eqno(B14)
$$

$$
\begin{array}{ll}
  \hat{B}f_5^{(5)}~=&\displaystyle
\frac{e^{-m_b^2/M_1^2-m_s^2/M_2^2}}{24M_1^8M_2^8}\{q^4({M_1}^2{M_2}^4 m_b m_s+{M_1}^4{M_2}^2(m_s^2-3{M_2}^2))\\
&-q^2({M_1}^2{M_2}^2 m_b^2 (m_s^2(2{M_1}^2+3{M_2}^2)-3(2{M_1}^2{M_2}^2+{M_2}^4))\\
&+{M_1}^2{M_2}^2 m_b m_s(m_s^2(3{M_1}^2+2{M_2}^2)-10{M_1}^2 {M_2}^2+{M_2}^4)\\
&+{M_2}^4 m_b^3 m_s(2{M_1}^2+{M_2}^2)+{M_1}^4(m_s^4({M_1}^2+2{M_2}^2)-2 m_s^2(3{M_1}^2{M_2}^2\\
&+2{M_2}^4)+2{M_2}^4 ({M_1}^2+3{M_2}^2)))-{M_2}^2 m_b^4({M_1}^2+{M_2}^2)(3{M_1}^2{M_2}^2\\
&-m_s^2({M_1}^2+2{M_2}^2))+{M_2}^4 m_b^5 m_s ({M_1}^2+{M_2}^2)+{M_2}^2 m_b^3 m_s(m_s^2(3{M_1}^4\\
&+4{M_1}^2{M_2}^2+{M_2}^4)-2{M_1}^2{M_2}^2(5{M_1}^2+3{M_2}^2))+{M_1}^2 m_b m_s (m_s^4 (2{M_1}^4\\
&+3{M_1}^2{M_2}^2+{M_2}^4)+m_s^2(-13{M_1}^4{M_2}^2-8{M_1}^2{M_2}^4+{M_2}^6)\\
&+4{M_1}^2{M_2}^4({M_1}^2+3 {M_2}^2))+m_b^2(-2 m_s^2 (3{M_1}^6{M_2}^2+5{M_1}^4{M_2}^4)\\
&+m_s^4({M_1}^6+4{M_1}^4{M_2}^2+3{M_1}^2{M_2}^4)+2{M_1}^4{M_2}^4({M_1}^2+3 {M_2}^2))\\
&+{M_1}^4 (m_s^6 ({M_1}^2+{M_2}^2)+2{M_2}^4 m_s^2({M_1}^2+{M_2}^2)-m_s^4(5{M_1}^2{M_2}^2\\
&+{M_2}^4)+4{M_1}^2{M_2}^6)\}
\times\langle \bar{s}s\rangle\; ,
 \end{array}
\eqno(B15)
$$

\newpage
$$
\begin{array}{ll}
 \hat{B}f_5^{(6)}~=&-\displaystyle
\frac{e^{-m_b^2/M_1^2-m_s^2/M_2^2}}{162M_1^8M_2^8(m_b^2-q^2)m_s^3}
\{{M_2}^4({M_1}^2+{M_2}^2) m_s^3 m_b^7+{M_2}^2({M_1}^4\\
&+3{M_2}^2{M_1}^2+2{M_2}^4) m_s^4 m_b^6+(({M_2}^6+4{M_1}^2{M_2}^4+3{M_1}^4 {M_2}^2) m_s^5\\
&-15{M_1}^2{M_2}^6 m_s^3) m_b^5+{M_1}^2(-54(-1+e^{\frac{m_s^2}{{M_2}^2}}){M_1}^4{M_2}^6+54{M_1}^4 m_s^2 {M_2}^4\\
&+({M_1}^4+4 {M_2}^2{M_1}^2+3{M_2}^4) m_s^6-(33{M_2}^6+14{M_1}^2{M_2}^4\\
&+2{M_1}^4{M_2}^2) m_s^4) m_b^4+{M_1}^2 m_s(-36 (-1+e^{\frac{m_s^2}{{M_2}^2}}) {M_1}^4{M_2}^6+(2{M_1}^4\\
&+3{M_2}^2{M_1}^2+{M_2}^4) m_s^6-(14{M_2}^6+40{M_1}^2{M_2}^4+7{M_1}^4{M_2}^2) m_s^4\\
&+2(10{M_1}^2{M_2}^6+51{M_1}^4 {M_2}^4) m_s^2) m_b^3+{M_1}^4 m_s^2(18(-1+e^{\frac{m_s^2}{{M_2}^2}}){M_1}^2{M_2}^6\\
&-({M_1}^2+10{M_2}^2) m_s^4 {M_2}^2+({M_1}^2+{M_2}^2) m_s^6+2(31{M_1}^2{M_2}^4\\
&-8{M_2}^6) m_s^2) m_b^2+72{M_1}^6{M_2}^6 m_s^3 m_b-18{M_1}^6{M_2}^6 m_s^4-{M_1}^2{M_2}^2q^6 m_s^3(m_s {M_1}^2\\
&+{M_2}^2 m_b)+q^4(-54(-1+e^{\frac{m_s^2}{{M_2}^2}}){M_2}^6{M_1}^6+54{M_2}^4 m_s^2{M_1}^6+{M_2}^2(3{M_1}^2\\
&+2{M_2}^2) m_b m_s^5 {M_1}^2+(3({M_1}^2+{M_2}^2) m_b^2-2({M_1}^4+5{M_2}^2{M_1}^2)){M_2}^2 m_s^4{M_1}^2\\
&+({M_1}^6+2{M_2}^2{M_1}^4) m_s^6+(({M_2}^6+3 {M_1}^2 {M_2}^4) m_b^3-14{M_1}^2{M_2}^6 m_b) m_s^3)\\
&+q^2(-{M_2}^4(3{M_1}^2+2{M_2}^2) m_s^3 m_b^5-{M_2}^2(3{M_1}^4+6{M_2}^2{M_1}^2+2{M_2}^4) m_s^4 m_b^4\\
&-(({M_2}^6+6{M_1}^2{M_2}^4+6{M_1}^4{M_2}^2) m_s^5-29{M_1}^2{M_2}^6 m_s^3) m_b^3+{M_1}^2(108(-1\\
&+e^{\frac{m_s^2}{{M_2}^2}}) {M_1}^4 {M_2}^6-108{M_1}^4 m_s^2{M_2}^4-(2{M_1}^4+6{M_2}^2{M_1}^2+3{M_2}^4) m_s^6\\
&+(33{M_2}^6+24{M_1}^2{M_2}^4+4{M_1}^4{M_2}^2) m_s^4) m_b^2-{M_1}^2 m_s(-36(-1+e^{\frac{m_s^2}{{M_2}^2}}){M_1}^4{M_2}^6\\
&+(2{M_1}^4+3{M_2}^2{M_1}^2+{M_2}^4) m_s^6-(14{M_2}^6+40{M_1}^2 {M_2}^4+7{M_1}^4{M_2}^2) m_s^4\\
&+2(10{M_1}^2{M_2}^6+51{M_1}^4{M_2}^4) m_s^2) m_b-{M_1}^4 m_s^2(18 (-1+e^{\frac{m_s^2}{{M_2}^2}}) {M_1}^2 {M_2}^6\\
&-({M_1}^2+10{M_2}^2) m_s^4{M_2}^2+({M_1}^2+{M_2}^2) m_s^6+2(31{M_1}^2{M_2}^4-8{M_2}^6) m_s^2))\}\\
&\times g^2\langle \bar{s}s\rangle^2\; .
\end{array}
\eqno(B16)
$$

\begin{center}{\bf Appendix C }\end{center}

The amplitudes in Eq.~(\ref{45}) and Eq.~(\ref{47}) are given here.

$$
\begin{array}{ll}
(\mathcal{\bar{A}}_1)_L=&\frac{G_F}{\sqrt{2}}\left[V_{cb}V_{cs}^*(C_1+\frac{C_2}{N_c})-V_{tb}V_{ts}^*(C_3+\frac{C_4}{N_c}
+C_5+\frac{C_6}{N_c}+C_7+\frac{C_8}{N_c}+C_9+\frac{C_{10}}{N_c})\right]h_0.
\end{array}
\eqno(C1)$$

$$
\begin{array}{ll}
(\mathcal{\bar{A}}_1)_\pm=
&\frac{G_F}{\sqrt{2}}\left[V_{cb}V_{cs}^*(C_1+\frac{C_2}{N_c})-V_{tb}V_{ts}^*(C_3+\frac{C_4}{N_c}
+C_5+\frac{C_6}{N_c}+C_7+\frac{C_8}{N_c}+C_9+\frac{C_{10}}{N_c})\right]h_\pm.
\end{array}
\eqno(C2)$$

$$
\begin{array}{ll}
(\mathcal{\bar{A}}_2)_L=\frac{G_F}{\sqrt{2}}C_1\left[V_{ub}V_{us}^*\frac{Q_u^2\alpha}{\pi}\int_{0}^{1}{2\bar{x}x(1+\ln{\frac{a^2}{\mu^2}})}dx+
V_{cb}V_{cs}^*\frac{Q_c^2\alpha}{\pi}\int_{0}^{1}{2\bar{x}x(1+\ln{\frac{b^2}{\mu^2}})}dx\right]h_0,
\end{array}
\eqno(C3)$$

$$
\begin{array}{ll}
(\mathcal{\bar{A}}_2)_\pm=\frac{G_F}{\sqrt{2}}C_1\left[V_{ub}V_{us}^*\frac{Q_u^2\alpha}{\pi}\int_{0}^{1}{2\bar{x}x(1+\ln{\frac{a^2}{\mu^2}})}dx+
V_{cb}V_{cs}^*\frac{Q_c^2\alpha}{\pi}\int_{0}^{1}{2\bar{x}x(1+\ln{\frac{b^2}{\mu^2}})}dx\right]h_\pm.
\end{array}
\eqno(C4)$$

$$
\begin{array}{ll}
(\mathcal{\bar{A}}_3)_L=\frac{G_F}{\sqrt{2}}C_1\left[V_{ub}V_{us}^*\frac{Q_u^2\alpha}{\pi}\int_{0}^{1}{2\bar{x}x\ln{\frac{a^2}{\mu^2}}}dx+
V_{cb}V_{cs}^*\frac{Q_c^2\alpha}{\pi}\int_{0}^{1}{2\bar{x}x\ln{\frac{b^2}{\mu^2}}}dx\right]h_0,
\end{array}
\eqno(C5)$$

$$
\begin{array}{ll}
(\mathcal{\bar{A}}_3)_\pm=\frac{G_F}{\sqrt{2}}C_1\left[V_{ub}V_{us}^*\frac{Q_u^2\alpha}{\pi}\int_{0}^{1}{2\bar{x}x\ln{\frac{a^2}{\mu^2}}}dx+
V_{cb}V_{cs}^*\frac{Q_c^2\alpha}{\pi}\int_{0}^{1}{2\bar{x}x\ln{\frac{b^2}{\mu^2}}}dx\right]h_\pm,
\end{array}
\eqno(C6)$$

$$
\begin{array}{ll}
(\mathcal{A}_1)_L=
&\frac{G_F}{\sqrt{2}}\left[V_{cb}^*V_{cs}(C_1+\frac{C_2}{N_c})-V_{tb}^*V_{ts}(C_3+\frac{C_4}{N_c}+C_5+\frac{C_6}{N_c}+C_7+\frac{C_8}{N_c}+C_9+\frac{C_{10}}{N_c})\right]h_0.
\end{array}
\eqno(C7)$$

$$
\begin{array}{ll}
(\mathcal{A}_1)_\pm=
&\frac{G_F}{\sqrt{2}}\left[V_{cb}^*V_{cs}(C_1+\frac{C_2}{N_c})-V_{tb}^*V_{ts}(C_3+\frac{C_4}{N_c}+C_5+\frac{C_6}{N_c}+C_7+\frac{C_8}{N_c}+C_9+\frac{C_{10}}{N_c})\right]h_\pm.
\end{array}
\eqno(C8)$$

$$
\begin{array}{ll}
(\mathcal{A}_2)_L=\frac{G_F}{\sqrt{2}}C_1\left[V_{ub}^*V_{us}\frac{Q_u^2\alpha}{\pi}\int_{0}^{1}{2\bar{x}x(1+\ln{\frac{a^2}{\mu^2}})}dx+
V_{cb}^*V_{cs}\frac{Q_c^2\alpha}{\pi}\int_{0}^{1}{2\bar{x}x(1+\ln{\frac{b^2}{\mu^2}})}dx\right]h_0,
\end{array}
\eqno(C9)$$

$$
\begin{array}{ll}
(\mathcal{A}_2)_\pm=\frac{G_F}{\sqrt{2}}C_1\left[V_{ub}^*V_{us}\frac{Q_u^2\alpha}{\pi}\int_{0}^{1}{2\bar{x}x(1+\ln{\frac{a^2}{\mu^2}})}dx+
V_{cb}^*V_{cs}\frac{Q_c^2\alpha}{\pi}\int_{0}^{1}{2\bar{x}x(1+\ln{\frac{b^2}{\mu^2}})}dx\right]h_\pm.
\end{array}
\eqno(C10)$$

$$
\begin{array}{ll}
(\mathcal{A}_3)_L=\frac{G_F}{\sqrt{2}}C_1\left[V_{ub}^*V_{us}\frac{Q_u^2\alpha}{\pi}\int_{0}^{1}{2\bar{x}x\ln{\frac{a^2}{\mu^2}}}dx+
V_{cb}^*V_{cs}\frac{Q_c^2\alpha}{\pi}\int_{0}^{1}{2\bar{x}x\ln{\frac{b^2}{\mu^2}}}dx\right]h_0,
\end{array}
\eqno(C11)$$

$$
\begin{array}{ll}
(\mathcal{A}_3)_\pm=\frac{G_F}{\sqrt{2}}C_1\left[V_{ub}^*V_{us}\frac{Q_u^2\alpha}{\pi}\int_{0}^{1}{2\bar{x}x\ln{\frac{a^2}{\mu^2}}}dx+
V_{cb}^*V_{cs}\frac{Q_c^2\alpha}{\pi}\int_{0}^{1}{2\bar{x}x\ln{\frac{b^2}{\mu^2}}}dx\right]h_\pm,
\end{array}
\eqno(C12)$$


\end{document}